\documentclass[apj,iop,revtex4]{emulateapj}
\usepackage{amsmath,hyperref,multirow, color}
\newcommand{\mytilde}{\raise.19ex\hbox{$\scriptstyle\sim$}}

\begin{document}

\title{Constraints on Cosmology and Baryonic Feedback with the Deep Lens Survey Using Galaxy-Galaxy and Galaxy-Mass Power Spectra}

\author{MIJIN YOON\altaffilmark{1}, M. JAMES JEE\altaffilmark{1,2}, \\ J. ANTHONY TYSON\altaffilmark{2}, SAMUEL SCHMIDT\altaffilmark{2}, DAVID WITTMAN\altaffilmark{2}, AND AMI CHOI\altaffilmark{3}}
\altaffiltext{1}{Department of Astronomy, Yonsei University, Yonsei-ro 50, Seoul, Korea; mjyoon@yonsei.ac.kr, mkjee@yonsei.ac.kr}
\altaffiltext{2}{Department of Physics, University of California, Davis, California, USA}
\altaffiltext{3}{Center for Cosmology and AstroParticle Physics, The Ohio State University, 191 West Woodruff Avenue, Columbus, OH 43210, USA}
\keywords{cosmological parameters --- gravitational lensing: weak ---
dark matter ---
cosmology: observations --- large-scale structure of Universe}

\begin{abstract}
We present cosmological parameter measurements from the Deep Lens Survey (DLS) using galaxy-mass and galaxy-galaxy power spectra in the multipole range $\ell=250\sim2000$. We measure galaxy-galaxy power spectra from two lens bins centered at $z\sim0.27$ and $0.54$ and galaxy-mass power spectra by cross-correlating the positions of galaxies in these two lens bins with galaxy shapes in two source bins centered at $z\sim0.64$ and $1.1$. We marginalize over a baryonic feedback process using a single-parameter representation and a sum of neutrino masses, as well as photometric redshift and shear calibration systematic uncertainties. For a flat $\Lambda$CDM cosmology, we determine $S_8\equiv\sigma_8\sqrt{\Omega_m/0.3}=0.810^{+0.039}_{-0.031}$, in good agreement with our previous DLS cosmic shear and the Planck Cosmic Microwave Background (CMB) measurements. Without the baryonic feedback marginalization, $S_8$ decreases by $\mytilde0.05$ because the dark matter-only power spectrum lacks the suppression at the highest $\ell$'s due to Active Galactic Nuclei (AGN) feedback. Together with the Planck CMB measurement, we constrain the baryonic feedback parameter to $A_{baryon}=1.07^{+0.31}_{-0.39}$, which suggests an interesting possibility that the actual AGN feedback might be stronger than the recipe used in the OWLS simulations. The interpretation is limited by the validity of the baryonic feedback simulation and the one-parameter representation of the effect.

\end{abstract}

\section{Introduction}
The initial conditions of our universe leave distinctive footprints on both the large scale structure and the cosmic expansion history. To determine these conditions (or more commonly referred to as cosmological parameters), a number of efforts have been made in the past few decades \citep[e.g.,][]{2003ApJS..148....1B, 2005ApJ...633..560E, 2011ARA&A..49..409A, 2012ApJ...746...85S,2017MNRAS.465.1454H} and projects with much greater survey powers will begin their operations in the current decade [e.g., Large Synoptic Survey Telescope (LSST)\footnote{\url{http://www.lsst.org}}; Wide-Field Infrared Survey Telescope (W-FIRST)\footnote{\url{https://wfirst.gsfc.nasa.gov}}; Euclid\footnote{\url{http://sci.esa.int/euclid}}; Square Kilometer Array (SKA)\footnote{\url{https://www.skatelescope.org}}; eROSITA\footnote{\url{http://www.mpe.mpg.de/eROSITA}}] through various observations including the cosmic microwave background (CMB), Type Ia supernovae, baryonic acoustic oscillations (BAO), galaxy clusters, and clustering properties of galaxies and dark matter. 

Studying clustering properties of galaxies and dark matter with weak lensing is among the most powerful methods among the aforementioned observations.  The weak-lensing signal is sensitive to both geometric and clustering properties of the universe. Past weak-lensing efforts have focused on measuring the clustering properties of the total mass \citep[e.g.,][]{2007MNRAS.376..771K,2010A&A...516A..63S,2012MNRAS.427..146H,Jee2013, 2014MNRAS.440.1322H, 2017MNRAS.465.1454H}. This so-called ``cosmic shear" measures shape correlations of distant galaxies to infer clustering properties of foreground total matter (dark matter + baryonic matter) without utilizing the information provided by intervening galaxies, the visible components of the foreground structure. The reason that the clustering properties of galaxies alone have not been used for precision cosmology is that galaxies are biased tracers of foreground structures. However, it has been suggested that this bias can be effectively constrained by combining galaxy auto-correlation and galaxy-mass correlation data \citep[e.g.,][]{Zhan2006, 2013MNRAS.430..767C, 2013MNRAS.432.1544M, 2017MNRAS.464.4045K, Abbott:2017wau, doi:10.1093/mnras/sty551}.
The combination enables cosmological parameter constraints because we can determine both the matter power spectrum $P_{\delta}$ and the galaxy bias $b$ via the relations $P^{gm}\propto b P_{\delta}$ and $P^{gg}\propto b^2 P_{\delta}$, where $P^{gm}$ and $P^{gg}$ are the galaxy-mass and galaxy-galaxy power spectra.
Hereafter, we will refer to this method based on the combined analysis of galaxy-galaxy and galaxy-mass correlations as G$^3$M for brevity; sometimes, the probe from the combination of all three two-point correlations (i.e., galaxy-galaxy, galaxy-mass, and mass-mass) is termed the ``$3\times2$pt" method. 

It is useful to probe the matter power spectrum of our universe through both cosmic shear and G$^3$M for the following reasons. First, as demonstrated by previous studies, the constraints from the G$^3$M method are nearly independent of those from cosmic shear even if the signals are extracted from the same survey data. Second, the two methods have different sensitivities to weak-lensing systematics. For example, the so-called additive shear bias tends to be cancelled in G$^3$M as tangential shears are azimuthally averaged around lens galaxies. On the other hand, in cosmic shear additive shear bias modulates the shear-shear correlation amplitude non-negligibly. Also, intrinsic alignments have much smaller impacts on the G$^3$M signals, where they become important only when a galaxy that is physically close to a lens is mistaken for a source by photometric redshift errors. However, in cosmic shear the shear-intrinsic ellipticity correlations (so-called GI systematics) affect the shear correlation between two galaxies separated by a large redshift difference.
Therefore, comparison of cosmological parameter constraints between the two methods provides critical insights on both instrumental and astrophysical systematics.

In this paper, we present cosmological  parameter  measurements  from  the  Deep  Lens  Survey  (DLS) by combining galaxy-mass and galaxy-galaxy power spectra. This is the third paper of the cosmology series from the DLS. In our two previous studies \citep{Jee2013,Jee2016}, we studied cosmology using two-dimensional (projected) and three-dimensional (tomographic) cosmic shear analyses. The previous DLS results are interesting in several aspects. First, despite the small survey area, the constraining power of the DLS is comparable or greater than those of other larger ($>100$ deg$^2$) surveys thanks to its depth.
Second, the results provide no tension with the Planck cosmological parameters based on CMB measurements \citep[hereafter Planck2015]{Ade:2015xua} while some recent weak-lensing results can be interpreted as indicating $>2\sigma$ tensions \citep[e.g.,][]{2015MNRAS.451.2877M, 2017MNRAS.467.3024L}. 
If ultimately found to be statistically significant, the discrepancy might be a serious challenge to the current $\Lambda$CDM paradigm. However, the conclusion should await scrutiny of all possible systematics. Occasionally, different analysis methods lead to non-negligible differences even for the same data \citep[e.g.,][]{Chang:2018rxd}. 
Certainly, this is one of the motivations of the current DLS study based on the G$^3$M method. Additionally, in the current study we
address the baryonic feedback effect using the power spectrum of \cite{2015MNRAS.454.1958M}, which models the power suppression on small scales due to AGN feedback. Therefore, the results from the current study serve as interesting comparisons to our previous cosmic shear-based results and also provide invaluable opportunity to reveal hidden systematics if the results are found to be statistically discrepant.

\begin{figure}[ht]\centering
\includegraphics[width = 0.475\textwidth]{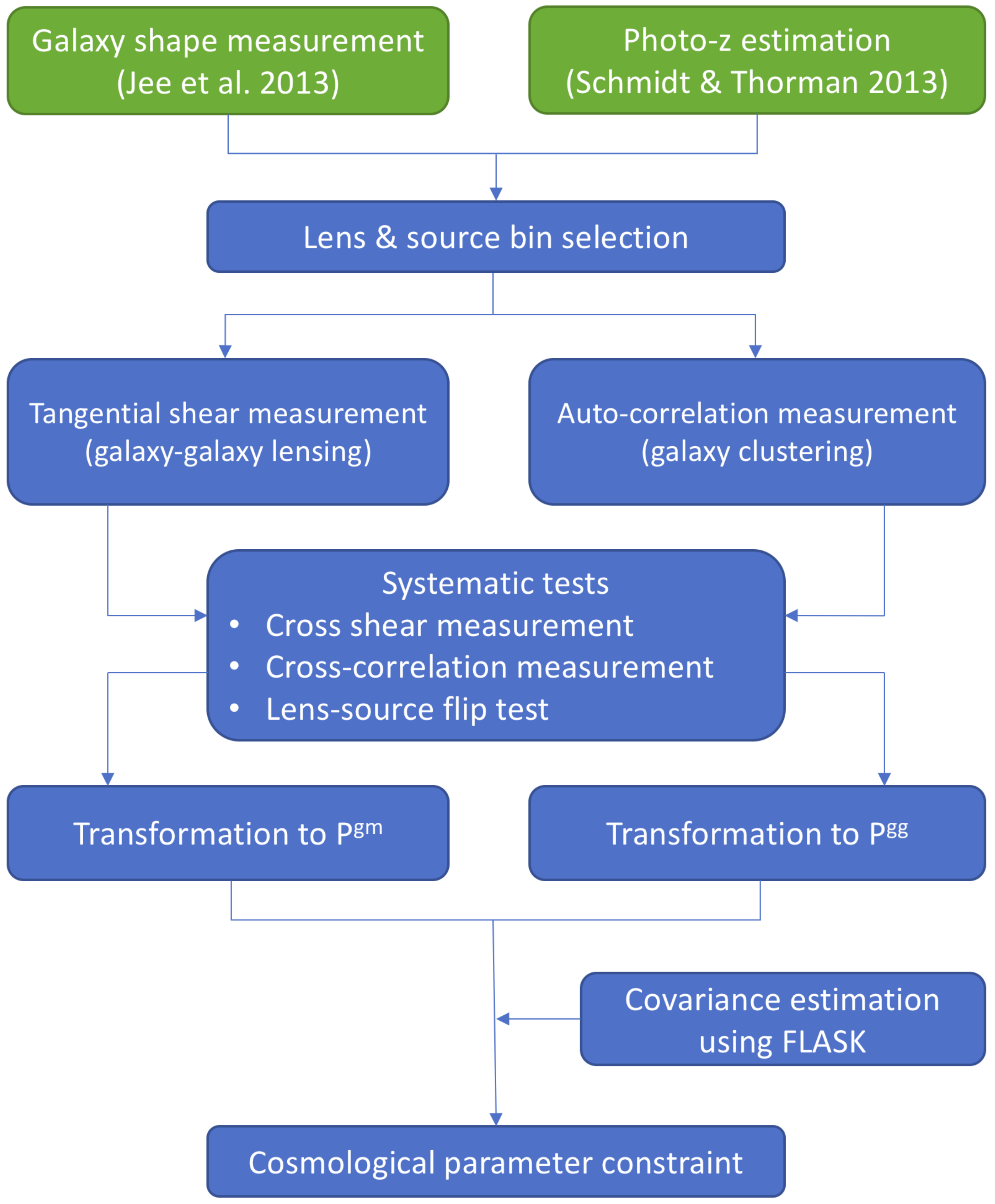}
\caption{Flowchart of our cosmological parameter constraints using galaxy-galaxy lensing and galaxy clustering signals from the DLS.}
\label{fig:flow_chart}
\end{figure}

Our paper is structured as follows. We present the theoretical background in \S\ref{sec:theory}. The DLS data and signal constructions are described in \S\ref{sec:data}. Our main cosmological parameter constraining results and discussions are presented in \S\ref{sec:results} and \S\ref{sec:discussion}, respectively before the conclusion in \S\ref{sec:conclusion}.
In Figure \ref{fig:flow_chart}, we summarize the flow of our analysis of the DLS data to constrain cosmological parameters from galaxy-galaxy lensing and galaxy clustering measurements.

\section{Theoretical Background} \label{sec:theory}

In the current study, we use power spectrum estimators to constrain cosmological parameters. The power spectrum estimators were first suggested in \cite{2002A&A...396....1S}. Among recent studies using combined analysis of galaxy-galaxy lensing and galaxy clustering, \citet{2016MNRAS.456.1508K,Kohlinger:2017sxk} and \cite{doi:10.1093/mnras/sty551} utilize power spectrum estimators, which have the following advantages. The power spectrum estimators are more fundamental than real space estimators because they are more directly related to the matter power spectrum whereas the correlation functions are obtained after
convolving these galaxy-galaxy/galaxy-mass power spectrum with highly oscillatory kernels.
Thus, if separation of small scales from large scales is clear in the matter power spectrum, in principle cosmological studies with power spectrum estimators can benefit from this scale separation.
 Moreover, the estimation of the power spectrum is computationally faster than the evaluation of its equivalent correlation function, which involves the aforementioned convolution and is only possible after the power spectrum is computed first. This gain in computational speed becomes an important factor when we sample a likelihood function numerous times in a high-dimensional parameter space.

Despite these advantages, most weak-lensing cosmological studies 
have been based on real-space correlation functions because the formal definition of the power spectrum involves integration angle from zero to infinity, which is not attainable in real observations. However, \cite{doi:10.1093/mnras/sty551} demonstrate that when they use band-limited power spectra, this weakness can be overcome. Below we summarize the formal definitions of the galaxy-mass and galaxy-galaxy power spectra and the corresponding band-limited power spectra used in the current analysis. 

\subsection{Galaxy-Mass Power Spectrum}

The projected galaxy-mass power spectrum $P^{gm}$ can be obtained from the matter power spectrum $P_\delta$ via the following relation:
\begin{multline}
    P^{gm}(\ell) = b\left (\frac{3H_0^2\Omega_m}{2c^2} \right)  \int_{0}^{\chi_H}d\chi \\ \times \frac{p_F(\chi)g(\chi)}{a(\chi)f_k(\chi)}P_\delta \left (\frac{\ell+1/2}{f_k(\chi)};\chi \right), \label{eqn_pgm_theory}
\end{multline}
\noindent
with $b$ the effective linear galaxy bias\footnote{In general, galaxy bias depends on scale or mass. However, here $b$ is the effective linear bias representing a collective value for the particular lens galaxy population.}, $H_0$ the Hubble constant, $\Omega_m$ the present matter density, $c$ the speed of light, $\chi$ the comoving distance, $\chi_H$ the comoving horizon distance, $a(\chi)$ the scale factor, $f_k$ the comoving angular diameter distance, $p_F(\chi)$ the redshift distribution of foreground galaxies, and $g(\chi)$ a lensing efficiency (geometric weight) factor
defined by 
\begin{equation} 
g(\chi) = \int_{\chi}^{\chi_H} d\chi' \, p_S(\chi')\frac{f_k(\chi'-\chi)}{f_k(\chi')}, 
\end{equation}
\noindent
where $p_S(\chi)$ is the source redshift distribution $p_S(\chi)d\chi=p_S(z)dz$. This galaxy-mass power spectrum $P^{gm}$ is also related to the mass-shear correlation (i.e., galaxy-galaxy lensing tangential shear) function $\gamma_T (\theta)$ via:
\begin{equation}
    P^{gm}(\ell) = 2\pi\int_{0}^{\infty}d\theta \,\theta \gamma_T(\theta)J_2(\ell \theta), \label{eqn_pgm_tan}
\end{equation}
\noindent
where $J_2$ is the 2nd order Bessel function of the first kind.

Since the evaluation of Equation~\ref{eqn_pgm_tan} requires our knowledge of $\gamma_t$ over angles from zero to infinity, Equations~\ref{eqn_pgm_tan} and \ref{eqn_pgm_theory} cannot be compared directly. Therefore, we use the following band power spectrum (as an estimator of $\ell^2 P^{gm}(\ell)$ for the $i^{th}$ $\ell$ interval):
\begin{eqnarray}
P_{band,i}^{gm} & = & \frac{1}{\Delta_i}\int_{\ell_{il}}^{\ell_{iu}}d\ell \, \ell P^{gm}(\ell) \label{eq:band_ps} \\
  & =  & \frac{2\pi}{\Delta_i}\int_{\theta_{min}}^{\theta_{max}}\frac{d\theta}{\theta} \gamma_T(\theta)[h(\ell_{iu}\theta) - h(\ell_{il}\theta)] \label{eq:tan2ps}
\end{eqnarray}
with 
\begin{equation}
h(x) = -xJ_1(x) -2J_0(x),  
\end{equation}
\noindent
where $J_{1(0)}$ is the first (zeroth) order Bessel function of the first kind,
and 
\begin{equation}
\Delta_i = ln(\ell_{iu}/\ell_{il}),
\end{equation}
\noindent
where $\ell_{iu}$ and $\ell_{il}$ are the upper and lower limits of the $i^{th}$ $\ell$ interval, respectively.

\subsection{Intrinsic Alignment Model}
\label{sec_IA_model}
The fundamental posit in weak-lensing is zero or negligible correlation of galaxy ellipticities in the absence of gravitational lensing. 
Certainly, this posit on intrinsic alignment (IA) becomes invalid in future surveys, where the interpretation is not limited by statistical errors. Cosmic shear studies from current precursor surveys have shown that although IA contamination causes a measurable shift in the best-fit parameter values, the amount of shift is still a small fraction of their statistical errors \citep[e.g.,][]{2013MNRAS.432.2433H,2017MNRAS.465.1454H,doi:10.1093/mnras/sty551}. In galaxy-galaxy lensing, systematic errors due to the IA contamination can arise when lens-source pairs are physically close;
large photometric redshift scatters make the lens-source separation imperfect.
As these source galaxies tend to align radially toward the lens galaxies, in principle the IA contamination in galaxy-galaxy lensing leads to signal suppression. In the current study, we estimate the level of signal suppression using
an IA model and find that the contamination is negligible and will not impact our cosmological parameter measurements as long as we avoid a lens-source pair whose redshift distributions do not overlap substantially.
Below, we present the details.

As in \cite{Jee2016}, we start with the following linear IA model of \cite{2001MNRAS.320L...7C} and \cite{2004PhRvD..70f3526H}:
\begin{equation}
    P_{\delta I} (z) = - A_{IA} ~C_1~ \rho_c \frac{\Omega_m}{D(z)} P_\delta (z), \label{eqn_IA}
\end{equation}
\noindent
where $\rho_c$ is the critical density of the universe today, $\Omega_m$ is the matter density today, $P_{\delta}(z)$ is the linear matter power spectrum at $z$, $A_{IA}$ is a dimensionless IA amplitude, $C_1$ is the coefficient fixed to the value $C_1=5\times10^{-14} h^{-2} M_{\sun}^{-1} \mbox{Mpc}^3$, and $D(z)$ is the growth factor at $z$ normalized to unity at $z=0$.
We replace the linear power spectrum $P_\delta$ in Equation~\ref{eqn_IA} with a non-linear version, following \cite{2007NJPh....9..444B}. 

Once the nonlinear IA power spectrum $P_{\delta I}$ is obtained, the corresponding IA contribution $P^{gI}$ to the galaxy-mass power spectrum $P^{gm}$ is estimated via:
\begin{equation}
    P^{gI} (\ell)= b \int_{0}^{\chi_H}d\chi \frac{p_F(\chi)p_S(\chi)}{f_k^2(\chi)}P_{\delta I}\left (\frac{\ell+1/2}{f_k(\chi)};\chi \right).
\end{equation}
\noindent
For the fiducial amplitude $A_{IA}=1$, we find that the fractional change in $P^{gm}$ is
$\sim3.7$\% at the most in our deliberate lens-source pair selection, which is much smaller than our statistical error. Considering the typical range of marginalization of the IA amplitude ($A_{IA}=-2\sim4$) in the literature, we conclude that the IA contamination is sub-dominant in our case. In this study, we marginalize over the amplitude of intrinsic alignment ($A_{IA}$) with a flat prior $[-4,4]$ for our main result. 
Note that the above IA power spectrum is slightly different from the one used in \cite{Jee2016}, where we also considered the luminosity-dependence using the \cite{2011A&A...527A..26J} measurement. The added sophistication would be superfluous here because of the negligible IA contribution.

\subsection{Galaxy Angular Power Spectrum}
\label{sec:ps_theory}
The galaxy angular power spectrum $P^{gg}$ of the lens galaxies is evaluated from the matter power spectrum $P_\delta$ as below: 
\begin{equation}
    P^{gg}(\ell) = b^2\int_{0}^{\chi_H}d\chi\, \frac{p_F^2(\chi)}{f_k^2(\chi)}P_\delta \left (\frac{\ell+1/2}{f_k(\chi)};\chi \right).
    \label{eq:pgg_theory}
\end{equation}
\noindent
This galaxy angular power spectrum $P^{gg}$ is related to the galaxy auto-correlation (often referred to as galaxy two-point correlation) function $w(\theta)$ through the following relation:
\begin{equation}
    P^{gg}(\ell) = 2\pi\int_{0}^{\infty}d\theta\, \theta w(\theta)J_0(\ell \theta).
\end{equation}
Analogously to the case of the galaxy-mass power spectrum $P^{gm}$,
we define the band-limited power spectrum for the galaxy angular power spectrum $P^{gg}$ as follows:
\begin{eqnarray}
P_{band,i}^{gg} & = & \frac{1}{\Delta_i}\int_{\ell_{il}}^{\ell_{iu}}d\ell \, \ell P^{gg}(\ell)  \label{eq:band_ps2}\\
  & = & \frac{2\pi}{\Delta_i}\int_{\theta_{min}}^{\theta_{max}}\frac{d\theta}{\theta} w(\theta)[q(\ell_{iu}\theta) - q(\ell_{il}\theta)]\label{eq:w2ps}
\end{eqnarray}
with 
\begin{equation}
q(x) = xJ_1(x).
\end{equation}

Note that the equalities between Equations~\ref{eq:band_ps} and \ref{eq:tan2ps} and between Equations~\ref{eq:band_ps2} and \ref{eq:w2ps} are not always valid. The equalities depend on the choice of the $\theta_{min}$ and $\theta_{max}$ values for the given $\ell$ ranges. The valid ranges of $\theta_{min}$ and $\theta_{max}$ were investigated in Appendix A of \cite{doi:10.1093/mnras/sty551}, who found that the estimate becomes slightly biased at the largest $\ell$ and applied corrections using theoretical predictions.
In our power spectrum estimation, we also address these issues (\S\ref{sec:ps_reconstruction}). 

\subsubsection{Power Spectrum and Baryonic Effects}
\label{sec:ps_and_baryon}
Robust evaluation of model power spectra $P^{gm}(\ell)$ and $P^{gg}(\ell)$ requires the accurate knowledge of the nonlinear matter power spectrum (Equations~\ref{eqn_pgm_theory} and \ref{eq:pgg_theory}). 
In our previous cosmic shear studies \citep{Jee2013,Jee2016}, we use the \cite{1998ApJ...496..605E} transfer function and the \cite{2003MNRAS.341.1311S} ``halofit" nonlinear power spectrum correction.
Experimenting with the \cite{2012ApJ...761..152T} version, which improved the accuracy of the \cite{2003MNRAS.341.1311S} power spectrum based on higher resolution N-body simulations, \cite{Jee2016} find that the $S_8 =\sigma_8 (\Omega_m/0.3)^{0.5}$ value decreases by $\mytilde$0.02, consistent with the findings of \cite{2015MNRAS.451.2877M}, who performed the comparison using the CFHTLenS lensing catalog. One weakness of the ``halofit" approach is that the result is valid only within a narrow range of cosmological parameters. 
\cite{2015MNRAS.454.1958M} overcame the limitation of the previous ``halofit" formalism with their modified version of the ``halo model." This new approach enables not only a significant reduction of the number of free parameters by more than a factor of three, but also a flexibility to accommodate a wider range of cosmological simulations with different initial conditions, which even include various baryonic effects. They show that it is possible for their revised halo model to describe varying degrees of baryonic effects with a single parameter using the relation between the two free parameters $\eta_0$ and $A_{baryon}$:
\begin{equation}
    \eta_0 = 0.98 - 0.12A_{baryon},  \label{eqn_baryon} 
\end{equation}
\noindent
where $\eta_0$ is a parameter characterizing $\eta$, which is referred to as the halo ``bloating" parameter in Mead et al. (2015). The parameter $A_{baryon}$ characterizes the relation between the concentration $C(M,z)$ of a halo with a mass $M$ at a redshift $z$ and its formation redshift $z_f$ via $C(M,z)=A_{baryon} (1+z_f)/(1+z)$.
The best fit values of $A_{baryon}$ are 3.13 for dark matter only simulation and 2.32 for the case with AGN feedback included.  Note that Equation~\ref{eqn_baryon} shows the updated result \citep{Joudaki:2017zdt} and is slightly different from the original relation published in Mead et al. (2015).
This flexibility of the \cite{2015MNRAS.454.1958M} approach provides an opportunity to investigate the impact of the baryonic physics on our power spectrum.
In our cosmological parameter estimation, we use the \cite{2015MNRAS.454.1958M} nonlinear power spectrum while marginalizing over the interval $2<A_{baryon}<4$, which brackets the AGN feedback result ($A_{baryon}=2.32$) from the OverWhelmingly  Large  cosmological  hydrodynamical  Simulations \citep[OWLS;][]{2010MNRAS.402.1536S, 2011MNRAS.415.3649V} and dark matter-only results ($A_{baryon}=3.13$). To implement this, we modified the \texttt{camb} and \texttt{pycamb} packages so that we can pass the $A_{baryon}$ parameter 
from \texttt{pycamb} to HMcode\footnote{\url{https://github.com/alexander-mead/HMcode}}, which computes the \cite{2015MNRAS.454.1958M} power spectrum. 

According to the current state-of-the-art cosmological hydro-simulations \citep[e.g.,][]{2011MNRAS.415.3649V, 2014MNRAS.444.1518V, 2014MNRAS.444.1453D, 2018MNRAS.475..676S}, AGN feedback suppresses the amplitude of the matter power spectrum substantially at $k>1~h~\mbox{Mpc}^{-1}$ \citep[e.g.,][]{2018arXiv180108559C}. 
When we examine the fractional change in $P^{gg}$ and $P^{gm}$ resulting from this matter power spectrum suppression corresponding to the OWLS simulation, 
we find that the effect is significant ($5-20$\%) across our entire ($250\lesssim \ell \lesssim 2000$) multipole range. We present the quantitative comparison in Appendix A.


\section{Data} \label{sec:data}


\subsection{Deep Lens Survey}

The DLS is composed of five widely separated fields (F1-F5). The two fields (F1 and F2) in the northern hemisphere were observed with Mosaic-1 at the NOAO/KPNO 4m Mayall Telescope and the three fields (F3, F4, and F5) in the southern hemisphere were observed with Mosaic-2 at the NOAO/CTIO 4m Blanco Telescope. The locations of the field centers are summarized in Table \ref{tab:field_loc}. The total survey area is $\mytilde20$ deg$^2$ with each field covering $\mytilde4$~deg$^2$ ($\mytilde2^{\circ}$ $\times$ 2$^{\circ}$). The DLS used more than $\mytilde120$ nights on these 20~deg$^2$ areas in order to reach down to $\mytilde26^{th}$ mag in $BVz^\prime$ bands and $\mytilde27^{th}$ mag in $R$ band (at the 5$\sigma$ level), approaching the depth of LSST. The depth enables us to obtain high-fidelity photometric redshifts and galaxy shears. The $R$ filter, where we measure galaxy shapes, was given priority whenever seeing is better than $\sim0\farcs9$. The mean cumulative exposure time in $R$ is $\mytilde$18,000~s per field while in other filters the exposure time is $\mytilde$12,000~s.   

We utilized the photo-$z$ data estimated with BPZ \citep{2004ApJS..150....1B} by \cite{2013MNRAS.431.2766S}, who calibrated the priors and the spectral energy distribution (SED) templates using $\mytilde$10,000 spectroscopic redshifts from the Smithsonian HEctospec Lensing Survey \citep[SHELS;][]{2005ApJ...635L.125G} on F2. The fidelity of the photo-$z$ estimations has been verified using an independent spectroscopic survey, PRIsm Multi-object Survey \citep[PRIMUS;][]{2011ApJ...741....8C} on F5 by \cite{2013MNRAS.431.2766S} and \cite{Jee2013}. 
The DLS galaxy shape catalog was obtained by fitting a PSF-convolved elliptical Gaussian to each galaxy image. The PSF was modelled by principal component analysis (PCA) method \citep{Jee&Tyson2011}. We refer readers to \cite{Jee2013} for shape measurement details.

\begin{deluxetable}{cccc}
\tablecaption{DLS Field Centers (J2000)}
\tablenum{1}
\tablehead{\colhead{Field} & \colhead{RA} &  \colhead{Dec} & \colhead{$l,b$}   }
\tablewidth{0pt}
\startdata
F1 & $00^h53^m25.3^s$ & +$12^{\circ}33'55''$ & $125^{\circ}, -50^{\circ}$\\[2pt]
F2 & $09^h19^m32.4^s$ & +$30^{\circ}00'00''$ & $197^{\circ}, 44^{\circ}$\\[2pt]
F3 & $05^h20^m00^s$ & -$49^{\circ}00'00''$ & $255^{\circ}, -35^{\circ}$\\[2pt]
F4 & $10^h52^m00^s$ & -$05^{\circ}00'00''$ & $257^{\circ}, 47^{\circ}$\\[2pt]
F5 & $13^h59^m20^s$ & -$11^{\circ}03'00''$ & $328^{\circ}, 49^{\circ}$
\enddata
\label{tab:field_loc}
\end{deluxetable}

\subsection{Lens and Source Selection}
\label{sec:lens_source_sel}
For measuring galaxy-galaxy lensing signals, we define two lens bins (L1, L2) and two source bins (S1, S2) over broad redshift ranges. To select galaxies in each bin, we use the peak redshift value ($z_b$) whereas for calculating the theoretical power spectrum, we stack the photometric redshift probability distribution $p(z)$ of individual galaxies (output by BPZ) to estimate the redshift distribution for each bin; a noticeable reduction of photo-$z$ bias when one uses $p(z)$ instead single-point estimates is shown in \cite{2009ApJ...700L.174W} and \cite{2013MNRAS.431.2766S}. 

The redshift range of our lens galaxies is $0.15<z_b<0.75$; we avoid galaxies whose $z_b$ is less than 0.15 because of a large discrepancy between photometric and spectroscopic redshifts in that low-redshift range \citep{Jee2013}. This lens redshift interval is divided into two lens bins: L1 ($0.15<z_b<0.4)$ and L2 ($0.4<z_b<0.75$). 
Although our using the stacked $p(z)$ curve instead of a collection of the $z_b$ values reduces the photo-$z$ bias, our detailed comparisons with the spectroscopic catalogs reveal that the mean redshift of the lens population in L1 would still be biased low by $\mytilde10$\% if left uncorrected whereas the bias would be negligible ($\mytilde1$\%) in L2 (see Appendix C). Therefore, in our cosmological parameter estimation we apply this $p(z)$ calibration to the L1 population in such a way that the means agree. We find that if this $p(z)$ calibration were omitted, our estimation of $S_8$ would be biased high by $\mytilde0.02$, which corresponds to $\mytilde50$\% of the statistical error.
 
We adopt the magnitude lower limit $m_R=18$ of \cite{Choi2012} for both lens bins while we use the upper limits $m_R$=21 and 22 for L1 and L2, respectively. Unlike \cite{Choi2012}, we, however, do not use absolute magnitudes as selection criteria because large photometric redshift scatters of individual galaxies can cause noise amplification. Stars are removed using the size-magnitude relation and shape criteria as described in \cite{Jee2013}.

Observing conditions such as depth variation, PSF, and extinctions can affect object selection and this can lead to non-negligible systematics \citep{2015MNRAS.454.3121M, leistedt2016}. For example, a large depth variation can leave measurable imprints on galaxy clustering signals.
Since our cutoff magnitudes are significantly brighter than the DLS limiting magnitude $\mytilde26.5$, the systematics due to depth variation is not a concern in our case.
Also, the DLS shapes are measured from co-added $R$-band images. Because we designed the survey in such a way that the $R$ filter gets priority whenever the seeing is better than $\mytilde0.9\arcsec$, the DLS seeing variation should not cause worrisome systematics.

We define source galaxies as follows. The redshift range is chosen to be $0.4<z_b<1.5$. The choice of the photo-$z$ upper limit is motivated by the DLS filters (BVRz$^{\prime}$) and the maximum redshift ($z\sim1.2$) of our photometric-spectroscopic redshift comparison sample. The interval is divided into two source bins: S1 ($0.4<z_b<0.75$) and S2 ($0.75<z_b<1.5$). 

The redshift range of the first source bin closely overlaps with that of the second lens bin. Therefore, 
the lensing signal ($P^{gm}$ measurement) from the L2-S1 pair is very weak and does not contribute to our cosmological parameter constraint. On the other hand, the overlapping $p(z)$ curves provide an opportunity to probe the intrinsic alignment. Because we regard the current intrinsic alignment model (Equation~\ref{eqn_IA}) as incomplete and desire to minimize the impact of this model incompleteness on our cosmology, we choose to exclude this L2-S1 pair in our main presentation of the cosmological parameter estimation. Nevertheless, we discuss our $A_{IA}$ measurement and cosmological parameter changes in case this L2-S1 pair is included.

The upper limit of the source magnitude is $m_R=24.5$, which corresponds to the approximate upper limit of the photometric-spectroscopic redshift comparison \citep{2013MNRAS.431.2766S}. 
According to the weak-lensing image simulation of \cite{Jee2013}, galaxies at $m_R>24.5$ require a large multiplicative factor in shear calibration. Therefore, applying this magnitude cut is our conservative measure to minimize the impact of our shear calibration and photo-$z$ uncertainties; we note that \cite{Jee2013,Jee2016} used source galaxies up to $m_R\sim26$. Because we measure shears from source galaxies, we also need to apply shape criteria. As in \cite{Jee2013,Jee2016}, we require 
the semi-minor axis of the best-fit (PSF-corrected) elliptical Gaussian to be larger than 0.4 pixels. In addition, we select sources whose ellipticity measurement error ($\sigma_{e,i}$) is less than 0.3. 

\begin{figure}
    \centering
    \includegraphics[width = 0.47\textwidth]{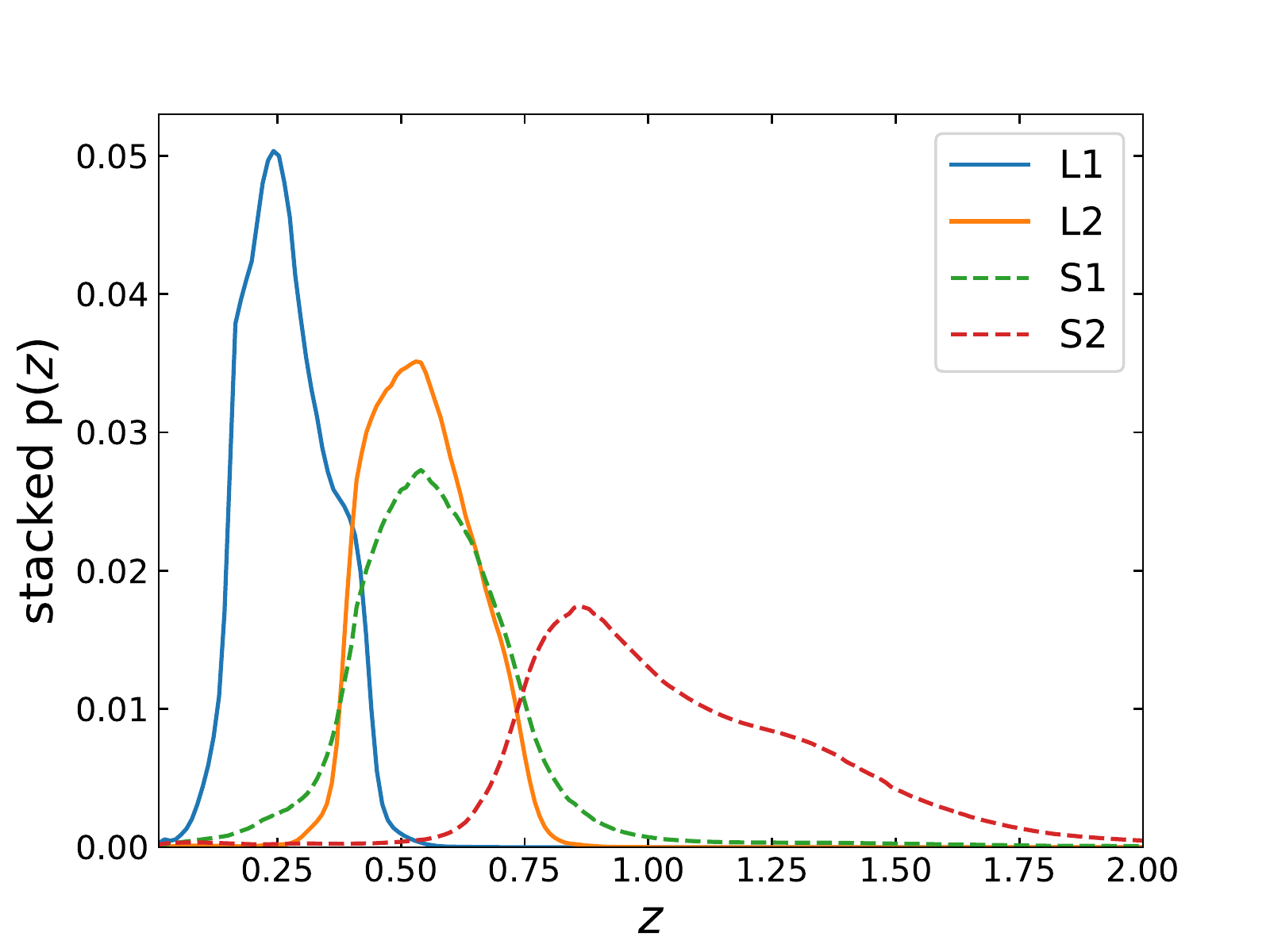}
    \caption{Redshift distributions of lens and source galaxies. The distribution in each bin is estimated by stacking the $p(z)$ curves of individual galaxies. We normalize the curves in such a way that their integrated areas are identical.}
    \label{fig:p_of_z}
\end{figure}

Table 2 summarizes our selection criteria and the resulting number of galaxies in each bin. The stacked redshift distribution of each bin is presented in Figure~\ref{fig:p_of_z}. 

\begin{deluxetable}{c  c  c  c  c  c  c r }
\tablecaption{Lens \& Source Selection}
\tablenum{2}
\tablehead{\multicolumn{2}{c}{\textbf{bins}} & \colhead{$\mathbf{z_b}^-$} &  \colhead{$\mathbf{z_b}^+$} &  \colhead{$\left< \mathbf{z} \right> $} & \colhead{$\mathbf{m_R}^-$ }&  \colhead{$\mathbf{m_R}^+$}& \colhead{ \textbf{\# of gal} } }
\tablewidth{0pt}
\startdata
\multirow{2}{*}{\textbf{Lens}} 
& L1 & 0.15 & 0.4& 0.270 & 18 &21 & 57,802\\
& L2 & 0.4 & 0.75 &0.542& 18 &22 & 98,267\\
\\[-6.5pt]
\hline
\\[-5.5pt]
\multirow{2}{*}{\textbf{Source}}
& S1 & 0.4 & 0.75 & 0.642& 21 &24.5 & 418,932\\
& S2 & 0.75 & 1.5 & 1.088 &21 &24.5 & 450,353
\enddata
\label{tab:lens_source_sel}
\end{deluxetable}

\subsection{Shear Calibration and Tangential Shear Measurement}

In general, weak-lensing shears are derived by measuring galaxy ellipticities and taking averages over  populations. A number of issues cause the average ellipticity to deviate from the true shear. Well-known difficulties include inaccurate point spread function (PSF) modeling, nonlinear relation between pixel noise and ellipticity (noise bias), discrepancy between galaxy model and real profiles (model bias), selection bias, incomplete de-blending, etc. 
Since the application of weak lensing to cosmology requires a sub-percent level accuracy in shear measurement, the community has invested significant efforts to develop and test various shear measurement techniques. The most prominent efforts include the public blind shear measurement challenge programs, in which  weak-lensing practitioners participate in analyzing and measuring weak-lensing shears from computer-generated galaxy images; the participants are blind to input shears. A variant of the DLS weak-lensing pipeline participated in the most recent public shear measurement challenge called the third GRavitational lEnsing Accuracy Testing \cite[GREAT3;][] {2015MNRAS.450.2963M} and won the challenge. 
The details of the galaxy shape measurement and shear calibration procedures are described in \cite{Jee2013}. Here we present the summary.

The DLS galaxy shapes are measured by fitting elliptical Gaussian profiles. The ellipticity $g$ is determined with the semi-major $\alpha$ and semi-minor $\beta$ axes using the relation $g=(\alpha-\beta)/(\alpha+\beta)$. 
The PSF effect is addressed by convolving the elliptical Gaussian with the model PSF prior to fitting. As mentioned above, the discrepancy between the Gaussian model and real galaxy profiles (model bias) is a non-negligible source of bias in shear estimation. Also, because of nonlinear coupling between pixel noise and shape parameter uncertainties, noise bias arises. 
\cite{Jee2013} address these shear calibration issues through image simulations \citep{Jee&Tyson2011} using 
real galaxy images from the Hubble Ultra Deep Field \citep[HUDF;][]{2006AJ....132.1729B}.
They determine the two shear calibration parameters in the following equation:
\begin{equation}
\gamma^{true} = m_{\gamma}\gamma^{obs} + C,
\end{equation}
\noindent
where $\gamma^{true}$ and $\gamma^{obs}$ are true and observed shears, respectively, $m_{\gamma}$ is a multiplicative correction parameter\footnote{Some authors prefer to use $m_{\gamma}^{\prime}=m_{\gamma}-1$ as the definition of the shear multiplicative bias.}, and $C$ is an additive correction parameter.
As the DLS additive correction is negligibly small ($\mytilde10^{-4}$), we only apply the multiplicative correction \citep{Jee2013}.
In principle, shear calibration is a function of many parameters such as PSF size, galaxy size and morphology, magnitude, noise level, etc. However, characterizing shear calibration with a large number of parameters is not feasible because of the limited number of galaxies in the HUDF; the result would be dominated by random fluctuations rather than by real trends. Therefore, we use the following single-parameter characterization:
\begin{equation}
m_\gamma = 6 \times 10^{-4}(m_R - 20)^{3.26} + 1.036,
\end{equation}
\noindent
where $m_R$ is the source magnitude.
This procedure is a good approximation because we conserve the HUDF galaxy properties such as size, morphology, etc. as a function of magnitude in our image simulations.  

After the application of the above shear calibration, we derive tangential shears as follows.
For each lens-source pair, a tangential shear is defined through:
\begin{equation}
g_T = -g_1 \cos 2\phi - g_2 \sin 2\phi\\
\end{equation}
\noindent
where $g_1$ and $g_2$ are the two components of the source galaxy ellipticity and $\phi$ is the position angle (measured counterclockwise) of the vector from the lens to the source with respect to a reference axis.

Obviously, a signal from a single lens-source pair is too small to detect, and thus it is necessary to stack signals over all lens-source pairs as follows:
\begin{equation}
\gamma_T^{raw}(\theta) = \frac{\sum\limits_{i,j} g_{T,ij} w_i}{\sum\limits_{i,j} w_i},
\end{equation}
\noindent
where $g_{T,ij}$ is the tangential shear of the $i^{th}$ source galaxy with respect to the $j^{th}$ lens galaxy, $\theta$ is
the distance between the lens-source pair, and $w_i$ is the inverse variance weight:
\begin{equation}
w_i = \frac{1}{\sigma_{e,i}^2+\sigma_{SN}^2}. \label{eqn_weight}
\end{equation}
\noindent
In Equation~\ref{eqn_weight}, $\sigma_{e,i}$ is the ellipticity measurement error for the $i^{th}$ source galaxy and $\sigma_{SN}$ is the ellipticity dispersion (shape noise) of the source population. 

Galaxy density fluctuations due to various masks and field boundaries increase the sample variance and also hamper the cancelling effects of residual additive biases in shear calibration through azimuthal averaging. To address the issue, we adopt the suggestion of \cite{Singh:2016edo} and subtract random catalog signals from the above raw tangential shears as follows:
\begin{equation}
\langle\gamma_T(\theta)\rangle = \langle \gamma_T^{raw} (\theta)\rangle - \langle\gamma_T^{random}(\theta)\rangle. 
\label{eq:tangental_shear}
\end{equation}
We find that this correction is important for tangential shear measurements on large scales ($\theta\gtrsim30\arcmin$) as shown in Appendix~D. We use the \texttt{Athena} code\footnote{\url{http://www.cosmostat.org/software/athena}} for measuring tangential shears and the \texttt{venice} code\footnote{\url{https://github.com/jcoupon/venice}} to generate the random points while taking care of star masks and field boundaries.

\begin{figure}
    \centering
    \includegraphics[width = 0.495\textwidth]{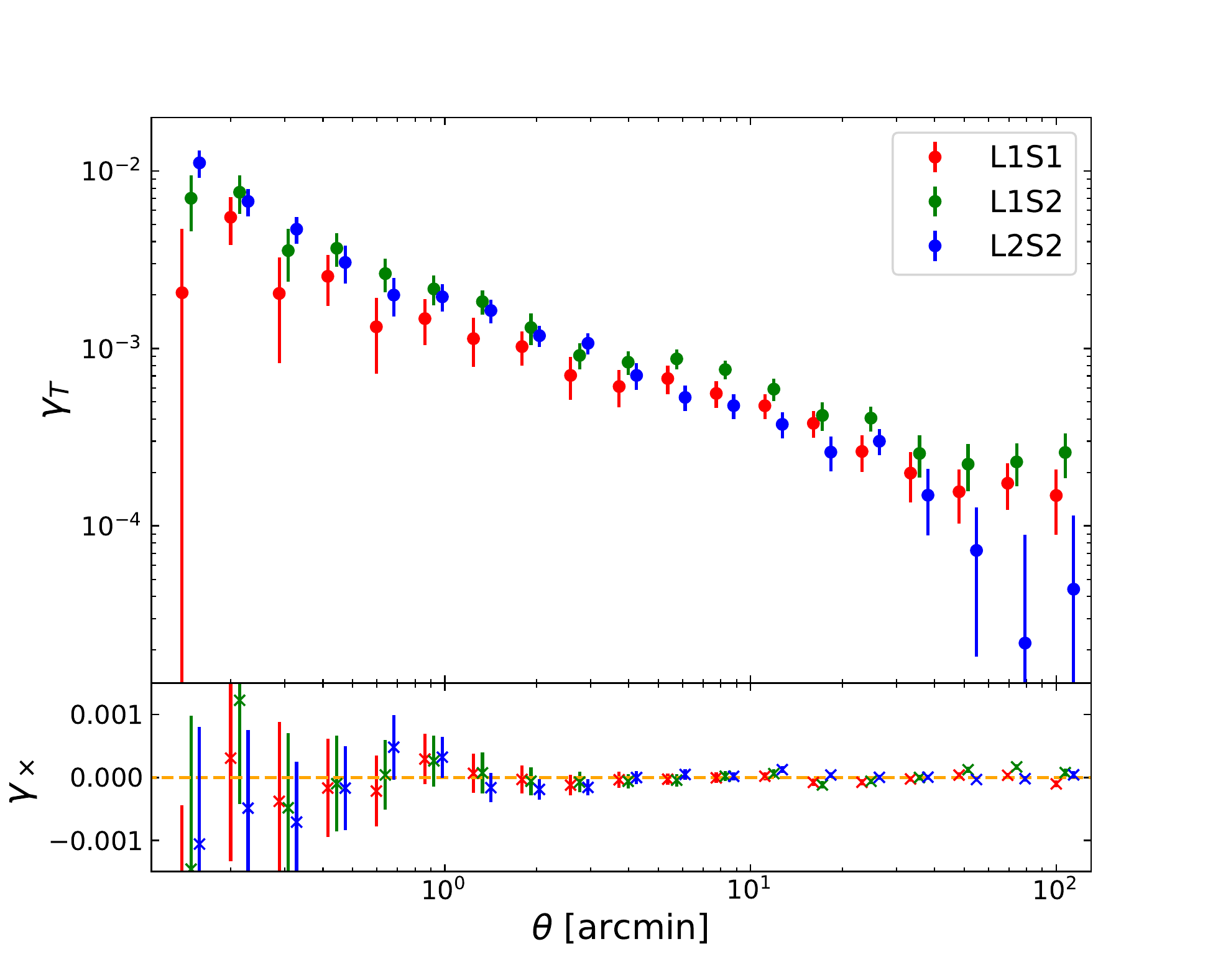}
    \caption{DLS shear measurement. Top: tangential shear profiles measured for the lens-source pairs, L1-S1, L1-S2 and L2-S2. Bottom: cross  (45$\degr$-rotated) shears measured to check residual systematics. The consistency of cross shears with zero shows that the shear additive errors are nicely controlled. The error bars estimated with our log-normal field simulations include the impact of shot noise (shape noise), field masks/boundaries, and the sample variance.}
    \label{fig:tangential_shear}
\end{figure}

We present our tangential shears for the L1-S1, L1-S2, and L2-S2 pairs in Figure~\ref{fig:tangential_shear}. 
As mentioned above, the displayed tangential shears are obtained after application of the random signal subtraction (Equation~\ref{eq:tangental_shear}).  The error bars are estimated with the log-normal field simulations (\S\ref{sec:cov_estimation}) and include the impact of shot noise (shape noise), field masks/boundaries, and the sample variance.  

To test residual lensing systematics, it is useful to examine cross shears, which are obtained by rotating source galaxy images by 45 degrees 
(the bottom panel of Figure~\ref{fig:tangential_shear}). As shown, they are consistent with zero on all scales for every lens-source bin pair, supporting the reliability of our tangential shear measurements.

\begin{figure}
    \centering
    \includegraphics[width = 0.495\textwidth]{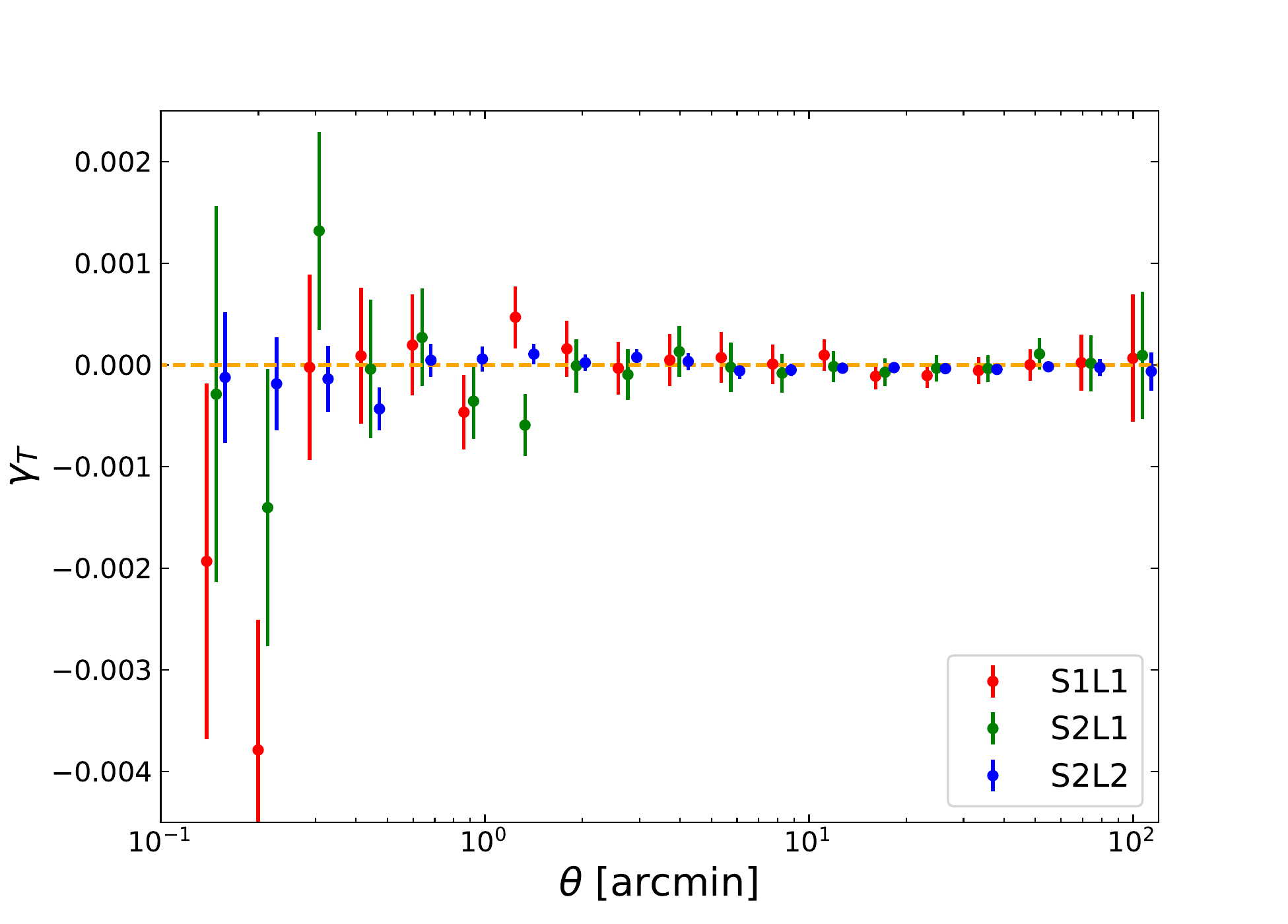}
    \caption{Lens-source flip test. The displayed signals are constructed by measuring tangential shears around source galaxies with lens galaxy shapes. Without the presence of measurable systematic errors in shear and photo-$z$ estimation, the resulting signals should vanish as shown. The errors are estimated with a 20$\times$20 block jackknifing in each field.}
    \label{fig:LS_flip_test}
\end{figure}

Another way to check residual systematics is the lens-source flip test, which examines the fidelity of both photo-$z$ estimation and shear measurement. In this test, lens and source bins are switched. That is, we measure tangential shears around source galaxies using lens galaxy shapes. If their redshift distributions indeed do not overlap significantly, as shown in Figure~\ref{fig:p_of_z}, the resulting signals should vanish. However, residual systematic errors in  photometric redshift and/or shear measurements would produce signals with non-zero amplitude. We perform this test for all three lens-source bin pairs, and the results are consistent with zero (Figure~\ref{fig:LS_flip_test}).   

\subsection{Galaxy Angular Correlation Measurement}
\label{sec:galaxy_angular_correlation}
The angular correlation function $w(\theta)$ is an excess probability of finding galaxies at a distance of $\theta$ with respect to that in a Poisson distribution:
\begin{equation}
 dP =  N [ 1+w(\theta) ] d\Omega ,
\end{equation}
\noindent
where $N$ is the mean number density of galaxies and $dP$ is the total expected number of galaxies at a distance $\theta$ within the solid angle $d\Omega$. If a galaxy bias is known, $w(\theta)$ alone can constrain cosmological parameters. In the current study, this galaxy bias is constrained by combining the galaxy clustering information with the galaxy-galaxy lensing signal.

In order to reduce systematic errors in the estimation of $w(\theta)$, a number of estimators have been suggested. We use the following estimator of \cite{1993ApJ...412...64L}:
\begin{equation}
w(\theta)=\frac{ \langle DD \rangle + \langle RR \rangle - 2 \langle DR \rangle } {\langle RR \rangle },
\end{equation}
\noindent
where $\langle DD \rangle$, $\langle DR \rangle$, and $\langle RR \rangle$ are the number of galaxy-galaxy pairs, galaxy-random pairs, and random-random pairs, respectively. 

When we blindly apply the above estimator to observational data with small areas, the amplitude of $w(\theta)$ is slightly underestimated by an additive factor known as the ``integral constraint". The deficit occurs because the average number of galaxies in the finite-size field becomes the reference to measure the excess probability. To correct for this bias, one should add the following constant \citep{1999MNRAS.307..703R}:
\begin{equation}
    IC = \frac{1}{\Omega^2}\int{w_{true}(\theta)}d\Omega_1d\Omega_2,
\label{eq:field_correction}
\end{equation}
\noindent
where $d\Omega_1$ and $d\Omega_2$ are two small patches within the observational field with the angular separation $\theta$, $w_{true}$ is the true angular correlation function, and the integral is evaluated over the entire observational field.
Although in the current study we assume the Planck2015 cosmology to calculate $w_{true}(\theta)$, we verify that the cosmology-dependence of the correction is negligibly small (\S\ref{sec:cosmology_dependence}).
The estimated IC values for L1 and L2 are 0.0126 and 0.0062, respectively. 

\begin{figure}
    \centering
    \includegraphics[width = 0.495\textwidth]{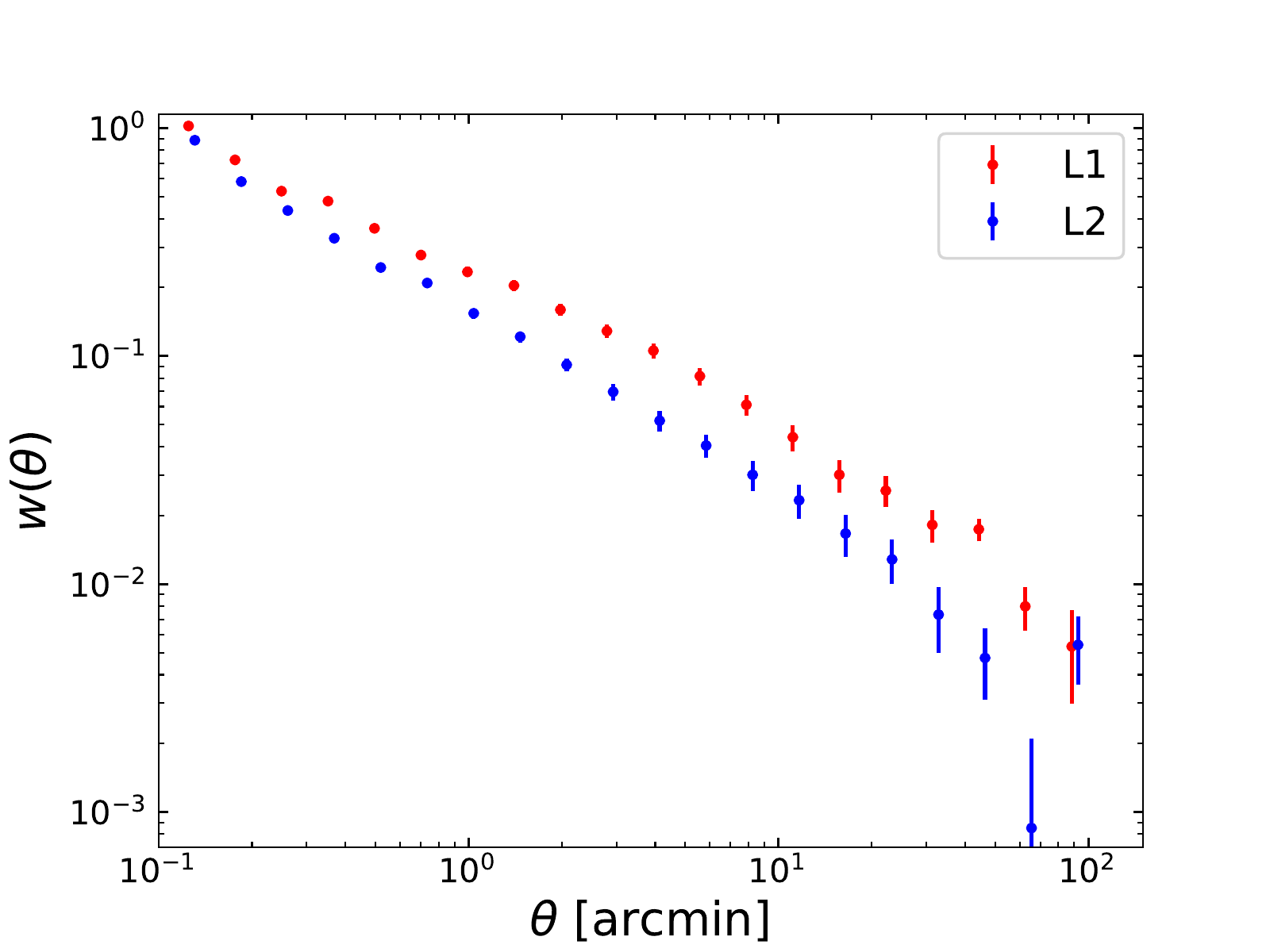}
    \caption{Galaxy angular correlation measured in the two lens bins L1 and L2. We use the \cite{1993ApJ...412...64L} estimator. The error bars are estimated using our log-normal field simulations, which include shot noise, field masking/boundaries, and the sample variance. IC corrections are included using Equation~\ref{eq:field_correction}.}
    \label{fig:w_of_theta}
\end{figure}

The galaxy angular correlations for the two lens bins L1 and L2 are plotted in Figure \ref{fig:w_of_theta}. As is done with the galaxy-shear correlations (\S\ref{sec:galaxy_angular_correlation}), the error bars are estimated using our log-normal field simulations, which
include various observational effects such as galaxy shot noise and field masks/boundaries, as well as the sample variance.
The galaxy angular correlations are measured in 20 logarithmic bins from $0\farcm1$ to 100$\arcmin$. The displayed correlation functions include the aforementioned integral constraints (Equation~\ref{eq:field_correction}).

\section{Results} \label{sec:results}

\subsection{Power Spectrum Reconstruction}
\label{sec:ps_reconstruction}
\begin{figure*}
    \centering
    \includegraphics[width = 0.495\textwidth, trim=0.0cm 0cm 0.5cm 0cm]{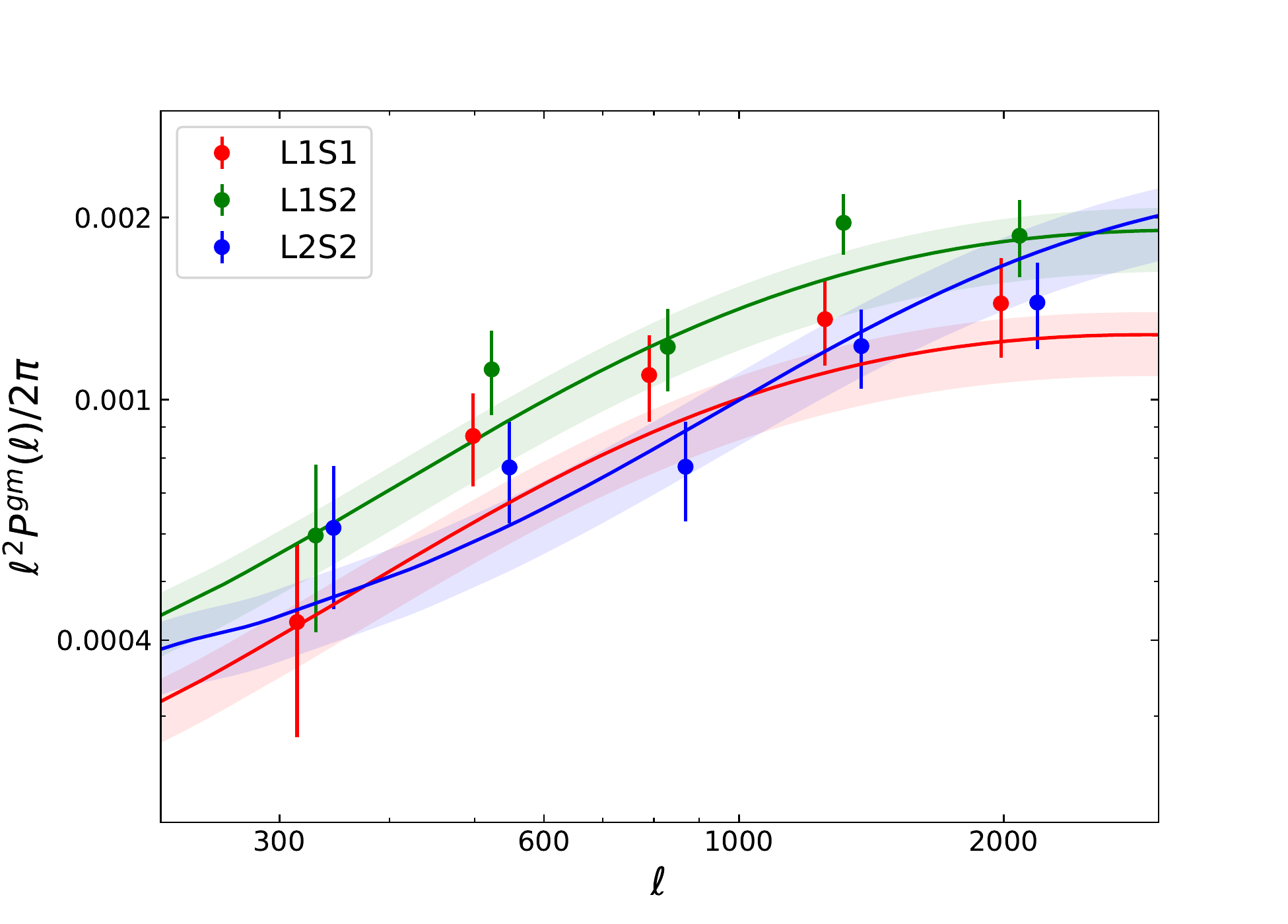}
    \includegraphics[width = 0.495\textwidth, trim=0.0cm 0cm 0.5cm 0cm]{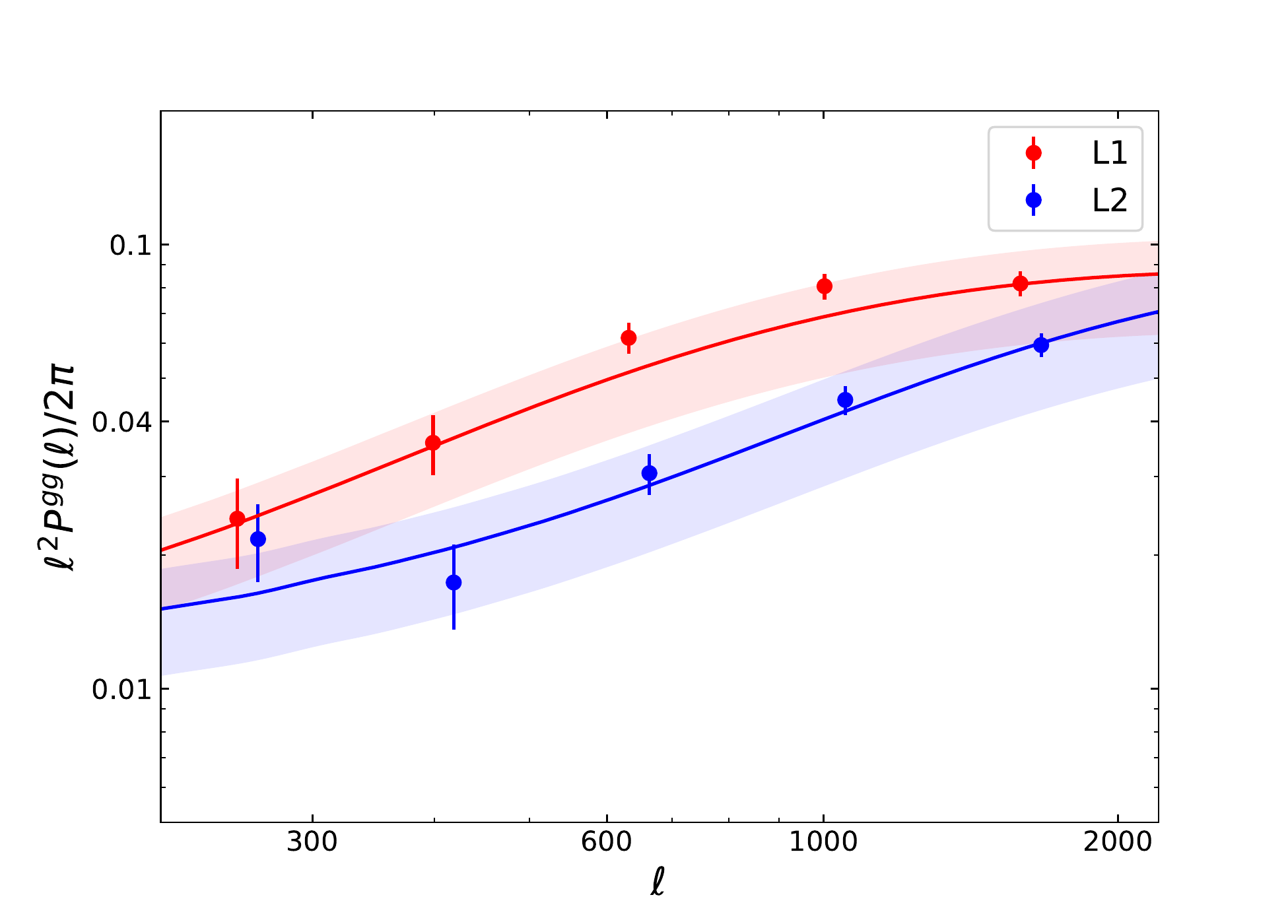}
    
    \caption{Observed DLS band power spectrum. left: galaxy-mass power spectrum $P^{gm}$ for the three lens-source pairs, L1-S1, L1-S2, and L2-S2. right: galaxy angular power spectrum for the two lens bins, L1 and L2. The $\ell$ bins are centered at $\ell$ = 251, 399, 632, 1002, and 1589 for $P^{gg}$ and $\ell$ = 314, 498, 790, 1252, and 1985 for $P^{gm}$. Slight horizontal shifts are applied to avoid clutter. Solid lines show theoretical predictions with the best-fit cosmological parameters. The error bars are  1-$\sigma$ ranges estimated using our log-normal field simulations. The offsets between the measurements and the best-fit lines are not statistically significant when we take into account of large uncertainty of galaxy biases $b$ whose 1-$\sigma$ levels are shown as shaded regions around the solid lines .}
    \label{fig:Pgm}
\end{figure*}

Following Equations~\ref{eq:tan2ps} and \ref{eq:w2ps}, we transform the tangential shear measurement $\gamma_T (\theta)$ to the band galaxy-mass power spectrum $P^{gm}_{band}(\ell)$ and the angular correlation function $w(\theta)$ to the band galaxy angular power spectrum $P^{gg}_{band}(\ell)$, respectively.

The range of the integral for $P^{gm}$ (Equation~\ref{eq:tan2ps}) is from 0.14$\arcmin$ to 100$\arcmin$ and the one
for $P^{gg}$ (Equation~\ref{eq:w2ps}) is from 0.12$\arcmin$ to 84$\arcmin$. The centers of the logarithmic $\ell$ bins are $\ell$ = 251, 399, 632, 1002, and 1589 for $P^{gg}$ and $\ell$ = 314, 498, 790, 1252, and 1985 for $P^{gm}$.
The different $\theta$ ranges and the corresponding $\ell$ ranges for $P^{gg}$ and $P^{gm}$ are deliberate choices. As  briefly mentioned in \S\ref{sec:ps_theory}, in order to accurately reconstruct a band power spectrum for each $\ell$ bin, \cite{doi:10.1093/mnras/sty551}   investigated valid $\theta$ ranges (see Appendix A of their paper). 
We repeat the experiment of \cite{doi:10.1093/mnras/sty551} using our DLS photometric redshifts and $\theta$ ranges and find that for the highest $\ell$ bin $P^{gm}$ allows $\theta_{min}$ as large as $\mytilde2\arcmin$, whereas $P^{gg}$ requires $\theta_{min}\lesssim0\farcm2$. 
On the other hand, the power spectrum evaluation at lowest $\ell$ bins requires the knowledge of $w(\theta)$ and  $\gamma_T(\theta)$ at large angles. In order to address the issue, we attach ``tails" of theoretically estimated $w(\theta)$ and  $\gamma_T(\theta)$ for the ranges from 85$\arcmin$ to 424$\arcmin$ and from 100$\arcmin$ to 493$\arcmin$, respectively. Here we use the Planck2015 cosmology for the computation. Although the exact values of the attached tails depend on cosmology, we verify that the impact of the assumed cosmology on our cosmological parameter determination is insignificant (\S\ref{sec:cosmology_dependence}).

The reconstructed power spectra are presented in Figure~\ref{fig:Pgm} with  1-$\sigma$ error bars. The solid lines (Equations~\ref{eqn_pgm_theory} and \ref{eq:pgg_theory}) show theoretical (including baryonic feedback effect and neutrino masses, \S\ref{sec:ps_and_baryon}, Appendix A and B) band power spectra computed at continuous $\ell$ ranges with the best-fit parameters (\S\ref{sec:param_result}).
The shaded regions represent the variations of the theoretical lines when the 1-$\sigma$ uncertainties of the two galaxy bias parameters $b_1$ and $b_2$ are considered.
In galaxy angular power spectrum, this uncertainty is magnified because the galaxy angular power spectrum is proportional to the galaxy bias squared ($P^{gg} \propto b^2 P_{\delta}$) whereas the galaxy-mass power spectrum is linear with the galaxy bias ($P^{gm} \propto b P_{\delta}$). 

For our likelihood evaluation (\S\ref{sec:likelihood}), we define a power spectrum data vector $\mathbf{p_d}$ with the reconstructed band power spectra using the following ordering:
\begin{equation}
\mathbf{p_d} = \left[P^{gg}_{L1}, P^{gg}_{L2}, P^{gm}_{L1S1},  P^{gm}_{L1S2},  P^{gm}_{L2S2} \right].
\end{equation}
\noindent
Since each band power spectrum is measured at five $\ell$ bins, the $\mathbf{p_d}$ vector is composed of 25 elements.

\subsection{Covariance Estimation}
\label{sec:cov_estimation}

In cosmological parameter estimation with a weak lensing survey, robust construction of a covariance matrix is paramount. The covariance should include the effects of galaxy shape noise, field maskings, boundary shapes, weak-lensing systematics, sample variances, etc. There have been a number of
suggestions for the estimation of the covariance matrix. In the linear regime, it is possible to derive the survey covariance analytically using Gaussian assumptions. The use of numerical simulations and ray-tracing methods has been a popular choice because the resulting covariance is valid in the nonlinear regime and one can easily incorporate observational features such as survey geometry, masking, etc. Another powerful method for producing high-fidelity covariances is to simulate weak-lensing galaxy catalogs using log-normal approximations, which is our choice in the current study. 

Approximation of the large scale structure of the universe with a log-normal distribution has been a popular choice \citep[e.g.,][]{1934ApJ....79....8H, 1980lssu.book.....P, 1991MNRAS.248....1C, 1993ApJ...403..450G,  2000MNRAS.314...92T,  2001ApJ...561...22K,2010MNRAS.409..355J, 2011A&A...536A..85H, 2014MNRAS.440...10A}. In this study, we use the Full-sky Lognormal Astro-fields Simulation Kit\footnote{\url{http://www.astro.iag.usp.br/~flask}} \citep[FLASK;][]{Xavier:2016elr}.
FLASK is useful for conveniently generating galaxy catalogs with a large sky coverage while including observational features. We use FLASK to estimate covariances for our galaxy-galaxy and galaxy-mass power spectra. 

\begin{figure}
    \centering
    \includegraphics[width = 8.5cm, trim=2.5cm 0cm 1.5cm 0cm, clip]{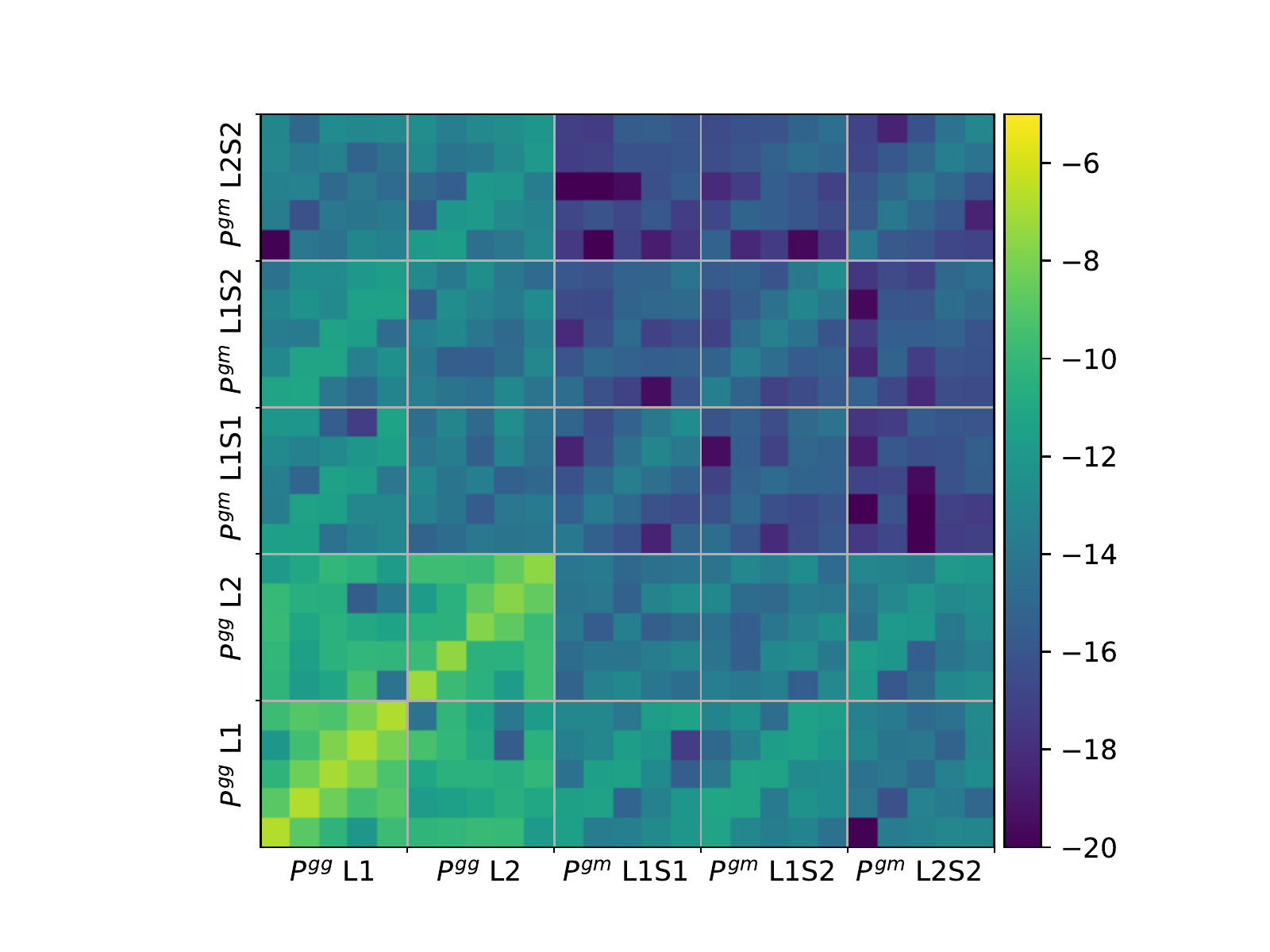}
    \caption{Covariance of the DLS band power spectrum. The covariance is estimated based on 100 FLASK simulations for each field (F1-F5). The ordering of the covariances are the galaxy angular power spectra $P^{gg}$ from L1 and L2 and the galaxy-mass power spectra $P^{gm}$ from the L1-S1, L1-S2, and L2-S2 pairs. Since each power spectrum has five $\ell$ bins, the total dimension of the covariance matrix is 25 by 25. The dominance of the (sub-)diagonal elements indicates that different scales are well separated; note that the level is depicted in a log scale.}
    \label{fig:FLASK_cov}
\end{figure}

FLASK generates catalogs that contain galaxy positions and shapes based on log-normal distributions, taking all combinations of galaxy-galaxy, galaxy-mass, and matter power spectrum between all lens and source bins as inputs. We produce 100 mock fields for each DLS field. To mimic the DLS observational features, we provide FLASK with the stacked photo-$z$ distributions for lenses and source bins, source density, galaxy shape dispersion, star masks, and field boundaries. 

The resulting FLASK catalogs are processed with the same analysis pipeline that is used for the DLS correlation function measurements in the current study and are converted to band power spectra. These power spectra from different realizations are combined to produce power spectrum covariances. The covariance matrix obtained in this way is shown in Figure~\ref{fig:FLASK_cov}. The ordering of the covariances are the galaxy-galaxy power spectra $P^{gg}$ from L1 and L2 and the galaxy-mass power spectra $P^{gm}$ from the L1-S1, L1-S2, and L2-S2 pairs. Since each power spectrum has five $\ell$ bins, the total dimension of the covariance matrix is 25$\times$25.
The dominance of the diagonal elements shows that the signals on different scales are only weakly correlated. 

The covariance matrix $\mathbf{C}$ obtained above can be utilized to quantify the {\it raw} signal-to-noise ratio (S/N), which is defined as:
\begin{equation}
\frac{S}{N}=\left( \mathbf{p_d}^{\intercal} \mathbf{C^{-1}} \mathbf{p_d}    \right )^{1/2}, \label{eqn:SNR}
\end{equation}
\noindent
where $\mathbf{p_d}$ is the data vector containing our observed band power spectra (\S\ref{sec:ps_reconstruction}). According to Equation~\ref{eqn:SNR}, the {\it raw} total S/N of our band power spectra from the DLS is 30.6. At face value this S/N estimate is higher than the one ($S/N=21.5$) for the DLS cosmic shear data presented in \cite{Jee2016}. However, we note that this larger S/N value does not directly translate to smaller parameter uncertainties because the two studies use different nuisance parameters (e.g., two galaxy bias parameters) and suffer from different degeneracies.



\subsection{Cosmological Parameter Constraints}

\subsubsection{Likelihood Sampling}
\label{sec:likelihood}

Our cosmological parameters are estimated by sampling the following likelihood function:
\small
\begin{equation}
\mathcal{L} = \frac{1}{(2\pi)^{n/2}|\mathbf{C}|^{1/2}}\exp \left [-\frac{1}{2}(\mathbf{p_d} - \mathbf{p_{th}})^{\intercal}\mathbf{C^{-1}}(\mathbf{p_d} - \mathbf{p_{th}}) \right ],  \label{eqn:likelihood}
\end{equation}
\noindent
where $\mathbf{p_{th}}$ is the theory vector predicted for a given set of cosmological parameters, $n$ is the number of elements in the vector, and $\mathbf{C}$ is the covariance matrix discussed in \S\ref{sec:cov_estimation}. Although the covariance depends on cosmology, it is treated as a constant in our parameter estimation (thus we ignore the determinant $|\mathbf{C}|$). We quantify the cosmology dependence in \S\ref{sec:discussion}.

One practical issue in deriving parameter constraints from the above likelihood is the sampling efficiency when the dimension is large and the likelihood function evaluation is computationally expensive. The traditional de facto standard tool is the Markov Chain Monte Carlo (MCMC) algorithm, which samples the likelihood in a high-dimensional parameter space based on a random walk. Thanks to increasing availability in parallelization, these time-consuming computations can be achieved within a reasonable amount of time. However, when one's interest is not only the inference (parameter value estimation), but also the model selection using Bayesian approach, one needs to compute Bayes factors, which require at least an order-of-magnitude more likelihood evaluations.

To overcome this computational challenge in Bayesian evidence estimation, one needs more efficient sampling algorithms than the traditional MCMC. In our study, we employ the nested sampling algorithm \citep{skilling2006}, which outputs
Bayesian evidence with much greater efficiency and provides parameter constraints as its byproducts. More specifically, we use the \texttt{multinest}\footnote{\url{https://github.com/JohannesBuchner/MultiNest}} package \citep{2009MNRAS.398.1601F}, which has been widely applied and tested in many cosmological studies such as \cite{Troxel:2017xyo}, \cite{Kohlinger:2017sxk}, \cite{2018arXiv180108559C}, etc.

\subsubsection{Prior Ranges}

\begin{deluxetable}{lcc}
\tabletypesize{\scriptsize}
\tablecaption{Prior Ranges Used in Cosmological Parameter Estimation}
\tablenum{3}
\tablehead{\colhead{parameters} & \multicolumn{2}{c}{prior range} }
\tablewidth{0pt}
\startdata
\\[-6.5pt]
\multicolumn{3}{c}{\textbf{Nuisance parameters}}\\
\\[-6.5pt]
photo-$z$ shift in L1, L2, S1, S2 ($\sigma_{zi}$), $\mathcal{N}$(0,0.02)  & -0.04 & 0.04 \\[3pt]
multiplicative shear error ($\sigma_{m_{\gamma}}$) & -0.03 & 0.03\\[3pt]
\hline
\\[-4.5pt]
\multicolumn{3}{c}{\textbf{Astrophysical parameters}}\\
\\[-6.5pt]
galaxy bias in L1 \& L2 ($b_i$)& 0.1 & 2.5\\[3pt]
baryon amplitude ($A_{baryon}$) & 2.0 & 4.0\\[3pt]
intrinsic alignment amplitude ($A_{IA}$) & -4.0 & 4.0\\[3pt]
\\[-6.5pt]
\hline
\\[-4.5pt]
\multicolumn{3}{c}{\textbf{Cosmological parameters}}\\ 
\\[-5.5pt]
matter density ($\Omega_{m}$) & 0.06 & 1.0\\[3pt]
baryon density ($\Omega_{b}$) & 0.03 & 0.06\\[3pt]
hubble parameter ($h$) & 0.55 & 0.85\\[3pt]
power spectrum normalization ($\sigma_8$) & 0.1 & 1.5\\[3pt]
spectral index ($n_s$) & 0.86 & 1.05\\[3pt]
sum of neutrino masses ($\Sigma_{\nu} m_\nu$/eV) & 0.06 & 0.9 \\[-6pt]
\enddata
\tablecomments{Displayed are the prior ranges of the 15 parameters used in our cosmological parameter estimation for the flat $\Lambda$CDM model (five nuisance, four astrophysical, and six cosmological parameters). Only photo-$z$ shifts employ Gaussian priors while others use flat priors.}
\label{tab:mcmc}
\end{deluxetable}

We define five nuisance parameters to address systematic uncertainties. To account for photo-$z$ systematic errors, we parameterize the photometric redshift probability of the lens and source redshift bins in the following way:
\begin{equation}
p_i(z) = p_i^o \left [(1+\sigma_{zi}) z \right],  \label{eqn_photo_z} 
\end{equation}
\noindent
where $p_i^o (z)$ is the observed (fixed) photometric redshift probability for the $i^{th}$ bin (derived from stacking the BPZ $p(z)$ curves of individual galaxies) and $p_i(z)$ is the randomized photometric redshift probability 
after the $z\rightarrow (1+\sigma_{z i})z$ mapping.
We let $\sigma_{zi}$ vary within the interval $[-0.04,0.04]$ following a zero-centered Gaussian distribution with a standard deviation of 0.02. This is based on the DLS photo-$z$ bias estimated by \cite{2013MNRAS.431.2766S}. Similar effects on parameter constraints are found when we instead applied $\pm3$\% flat prior employed in \cite{Jee2013,Jee2016}; this 3\% flat prior was motivated by the 3\% difference measured in photometric redshift comparison between the  VIMOS-VLT Deep Survey and the Hubble Deep Field North (HDF-N) priors. Since we have two redshifts bins for both lens and source, the total number of the $\sigma_{zi}$ parameters is four.

When marginalizing over the multiplicative shear calibration bias in $m_{\gamma}$, we assume a 3\% flat prior as in our cosmic shear studies \citep{Jee2013,Jee2016}. We modify the model power spectra as the following:
\begin{equation}
P^{gm \prime}(\ell) = (1+\sigma_{m_\gamma})P^{gm}(\ell).
\end{equation}
This marginalization is equivalent to the covariance correction used in \cite{Troxel:2018qll}.

For the two galaxy bias parameters $b_1$ and $b_2$, we apply a flat prior ranging from 0.1 to 2.5. These two parameters are highly degenerate with $\Omega_m$ and $\sigma_8$ and thus require sufficiently large intervals to minimize parameter estimation bias imposed by the prior interval. We verify that both bias parameters are constrained well within this prior interval and enlarging it further does not change our results.

Our main cosmological parameter constraints are obtained for a flat $\Lambda$CDM universe with baryonic feedback. For $\Omega_m$ and $\sigma_8$, we use the prior intervals $[0.06,1.0]$ and $[0.1,1.5]$, respectively. Similarly to $b_1$ and $b_2$, these two cosmological parameters are well constrained within the prior ranges. 
For the Hubble constant, we use the interval $[0.55,0.85]$, which brackets the 4-$\sigma$ lower- and upper-limits of both the Planck and direct measurements \citep{Ade:2015xua, 2018ApJ...855..136R}.
Our prior for the scalar spectral index $n_s$ varies with a uniform probability within $[0.86,1.05]$, again well encompassing the current constraints. 

As mentioned in \S\ref{sec_IA_model}, we choose to marginalize over $-4<A_{IA}<4$ to address the IA contamination. Because we carefully select the lens-source pairs in such a way that the IA contamination is minimized, the inclusion of this IA model does not produce significantly different results from those obtained without it. Also, enlarging the interval to $-6<A_{IA}<6$ yields only negligible changes in our parameter estimation (Appendix F). 

Currently, the most uncertain model parameter is $A_{baryon}$. According to \cite{2015MNRAS.454.1958M} who base their analysis on the OWLS results, the power spectrum from the dark matter-only simulation corresponds to $A_{baryon}=3.13$\footnote{This value is derived for the simulation using the cosmological parameters favored by the Planck CMB data. The exact value depends on the assumed cosmology. For example, when the simulation uses WMAP3 cosmological parameters, the best-fit value becomes $A_{baryon}=3.43$.}. For a simulation that has prescriptions for baryonic physics such as gas cooling, heating, star formation and evolution, chemical enrichment, and supernovae feedback (however without AGN feedback), the preferred value slightly increases to $A_{baryon}=3.91$, which nevertheless is not a statistically meaningful difference from the former case. 
A significant change occurs when the prescription includes AGN feedback, which reduces $A_{baryon}$ to 2.32. 
Since currently there is no consensus on the exact impact of baryonic feedback \citep[e.g.,][]{2018arXiv180108559C}, we rely on the \cite{2015MNRAS.454.1958M} results based on the OWLS simulations and choose $A_{baryon}$ to vary within the interval $[2,4]$, which brackets both the dark matter only simulation case ($A_{baryon}=3.13$) and the largest departure from it ($A_{baryon}=2.32$). Note that the same interval $[2,4]$ is also used in \cite{2017MNRAS.465.1454H} and \cite{2017MNRAS.465.2033J}. As discussed \S\ref{sec:baryon_model_sel}, only an upper bound for $A_{baryon}$ is constrained within this $[2,4]$ range and it is necessary to enlarge this prior range to $[0.1,4]$ in order to obtain a meaningful constraint on $A_{baryon}$. Nevertheless, for our main presentation, we report the results from the original $[2,4]$ interval because the validity of the single-parameter representation has not been tested at $A_{baryon}\lesssim2$.

Finally, we marginalize over a sum of neutrino masses ($\Sigma_{\nu} m_{\nu}$) within the flat prior range [0.06~eV, 0.9~eV]. 
The theoretical lower limit is $\mytilde0.06$ eV for standard-model active neutrinos with the normal hierarchy. According to Planck2015, the upper limit of the 95\% confidence regions for the sum of neutrino masses varies from $\mytilde0.2$~eV to $\mytilde0.7$~eV depending on the combinations of the Planck power spectra, Planck lensing, and external data. Thus, our prior range
[0.06~eV, 0.9~eV] amply brackets the current theoretical and observational constraints.
Including neutrino masses are important because
similarly to baryonic feedback, massive neutrinos also suppress the amplitude of the power spectrum on small scales (Appendix B). For simplification, we consider the case with one massive neutrino and two zero-mass neutrino species.

For our baseline cosmology (flat $\Lambda$CDM), the total number of parameters is 15 (five nuisance, four astrophysical, and six cosmological parameters). We summarize their prior ranges in Table 3. In Appendix F, we discuss the impact of our prior choices on parameter constraints.

\begin{figure*}
    \centering
    \includegraphics[width = 0.95\textwidth]{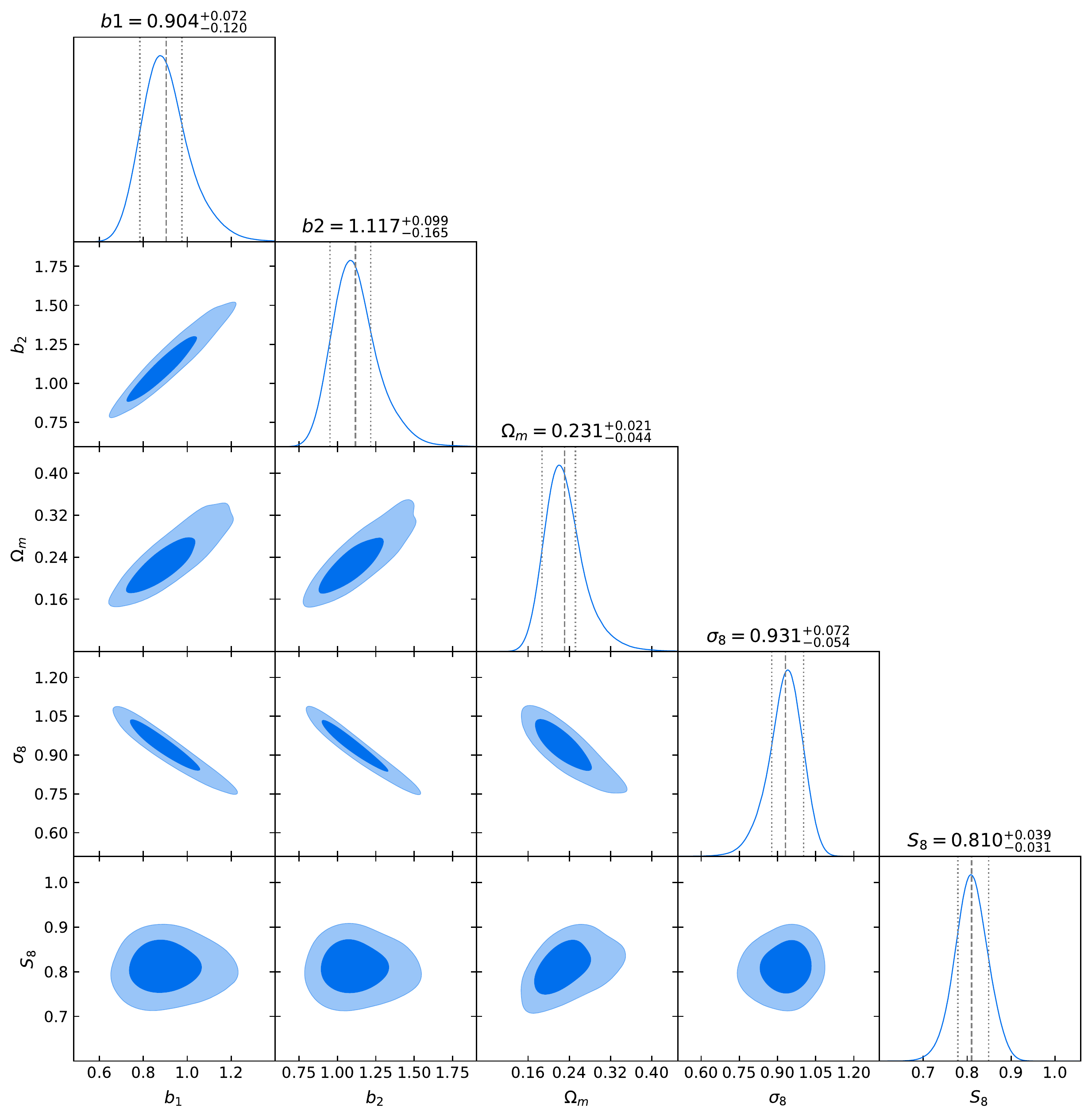}
    \caption{Constraints on $b_1$, $b_2$, $\Omega_m$, and $\sigma_8$ estimated by our nested sampling for the flat $\Lambda$CDM model. Also displayed is the constraint on the derived parameter $S_8\equiv \sigma_8 (\Omega_m/0.3)^{0.5}$. The three parameters $b_1$, $b_2$, and $\Omega_m$ are positively correlated with one another while $\sigma_8$ is anti-correlated with these three parameters. $S_8$ shows no significant correlation with the other cosmological parameters. }
    \label{fig:triangle}
\end{figure*} 

\subsubsection{Parameter Estimation Results}
\label{sec:param_result}

Figure~\ref{fig:triangle} displays our parameter constraint results. The 
constrained parameters are the matter density ($\Omega_m$), normalization ($\sigma_8)$, and two effective galaxy bias parameters $b_1$ and $b_2$; the $S_8\equiv \sigma_8 (\Omega_m/0.3)^{0.5}$ parameter is not independent.
It is clear that those four constrained parameters are highly degenerate with one another. The degeneracy between $\Omega_m$ and $\sigma_8$ arises because the overall power spectrum amplitude $A_p$ measured by weak lensing is $A_p \sim \sigma_8 \Omega_m^{\alpha}$, where the exponent $\alpha$ is $\mytilde0.5$. This motivates the definition of $S_8\equiv \sigma_8 (\Omega_m/0.3)^{0.5}$, which is useful when results from different studies are compared. The current DLS G$^3$M constraint with baryonic feedback marginalization is $S_8=0.810^{+0.039}_{-0.031}$ while without the marginalization we obtain a lower value $S_8=0.753^{+0.040}_{-0.030}$. The shift in $S_8$ arises from the power spectrum suppression due to the baryonic feedback (Appendix A).

\begin{figure*}
    \centering
    \includegraphics[width = 0.9\textwidth]{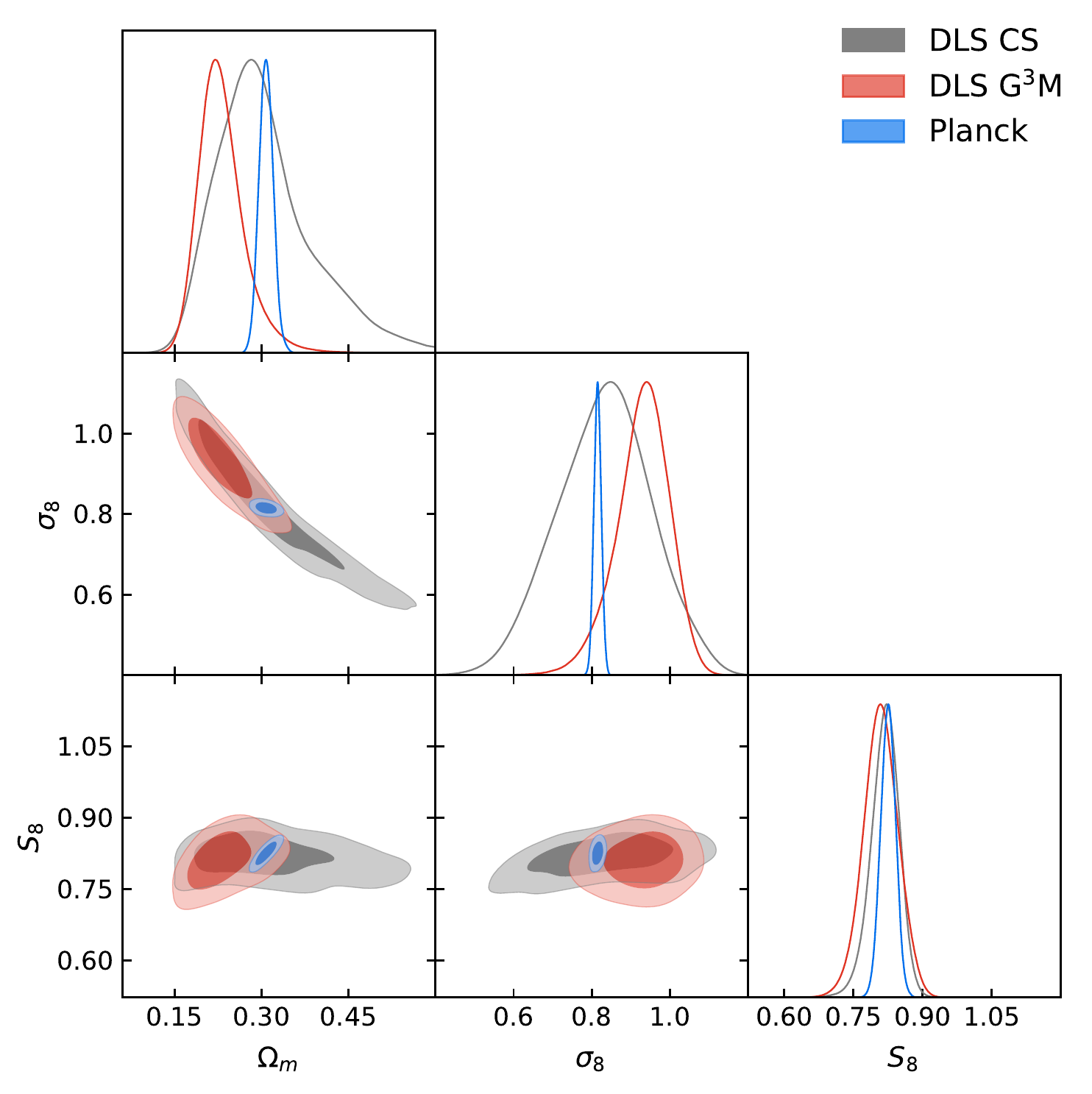}
    \caption{Constraints on $\sigma_8$, $\Omega_m$, and $S_8$ for a flat $\Lambda$CDM model from the G$^3$M method (red), from cosmic shear (grey), and from Planck2015 (blue). A smoothing kernel is applied. Two weak lensing analyses in DLS agree well and show no tension with Planck.}
    \label{fig:sigma8_omegam_contour}
    
\end{figure*}

The DLS G$^3$M constraint is in good agreement with the value derived from the DLS tomographic cosmic shear $S_8=0.818_{-0.026}^{+0.034}$ \citep{Jee2016} as shown in Figure~\ref{fig:sigma8_omegam_contour}. Since the DLS cosmic shear result was obtained without marginalizing over the baryonic effect parameter, we expect that the $S_8$ value would increase slightly when we repeat the analysis with the same marginalization, which is the subject of a future investigation. The $S_8$ uncertainty from the cosmic shear is $\mytilde86$\% of the G$^3$M uncertainty. However, this $S_8$ uncertainty alone should not be used to judge the overall S/N of the DLS G$^3$M data because $S_8$ is a measure of the parameter constraints in a particular projection and its uncertainty is proportional to the width of the $\Omega_m$-$\sigma_8$ ``banana". One way to compare the $\Omega_m$-$\sigma_8$ degeneracy breaking power (i.e., reducing the length of  the ``banana") is simply to compare the uncertainties of the marginalized parameter constraints.
The $\Omega_m$ uncertainty from the G$^3$M analysis is 
$\mytilde56$\% of the cosmic shear result. The resulting shrinkage of the area within the 1-$\sigma$ contour (Figure~\ref{fig:sigma8_omegam_contour}) shows that the information content of the G$^3$M signal is greater, as also indicated by the raw S/N comparison (\S\ref{sec:cov_estimation}).
We defer detailed comparison of our $S_8$ measurement with those from other studies to \S\ref{sec:discussion}.

Since the galaxy bias parameters are degenerate with the amplitudes of the galaxy-galaxy and galaxy-mass power spectra ($P^{gg}\sim b^2 P_{\delta}$ and $P^{gm}\sim b P_{\delta}$), the tight correlations (degeneracies) of these parameter values with $\Omega_m$ and $\sigma_8$ are expected as shown in Figure~\ref{fig:triangle}. The marginalized $b_1$ (L1) and $b_2$ (L2) parameter constraints are $0.904^{+0.072}_{-0.120}$ and $1.117^{+0.099}_{-0.165}$, respectively. Since the error bars of the two parameters marginally overlap, the difference in their central values is only a weak indication of the possible bias evolution from $\left < z \right >\sim 0.54$ to $\mytilde0.27$.

\begin{figure}
    \centering
    \includegraphics[width = 0.495\textwidth]{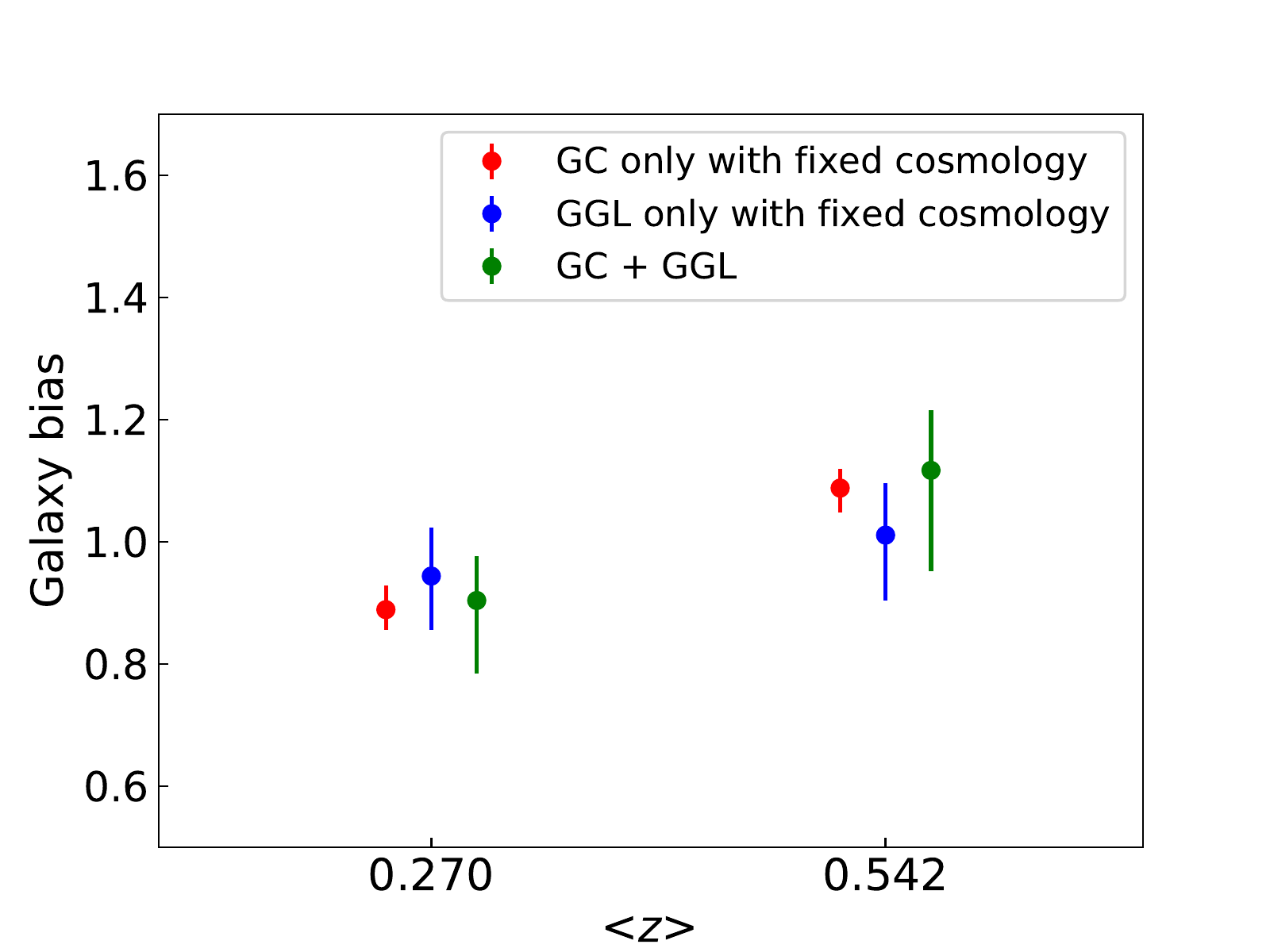}
    \caption{Galaxy bias measured for each lens bin under different conditions: galaxy clustering ($P^{gg}$) only with fixed cosmology (best-fitted $\Omega_m$, $\sigma_8$), galaxy-galaxy lensing ($P^{gm}$) only with fixed cosmology, and both with free cosmological parameters. Independently measured galaxy biases are consistent with each other, validating the combined analysis of galaxy clustering and galaxy-galaxy lensing}
    \label{fig:bias_test}
\end{figure}

One useful consistency test between the galaxy-mass and the galaxy auto power spectra is to constrain galaxy biases independently from each method for a fixed cosmology. Figure~\ref{fig:bias_test} displays the results for the following two cases. In case 1, we fix $\Omega_m$ and $\sigma_8$ to our best-fit values and use only the galaxy clustering signal (GC-only). In case 2, we again fix $\Omega_m$ and $\sigma_8$ to our best-fit values, but this time use only the galaxy-mass  power spectrum data (GGL-only). The results from these two cases are consistent with the ones that we obtain after marginalizing over cosmological parameters. This test supports the internal consistency of the DLS data. 

\begin{figure}[h]
    \centering
    \includegraphics[width = 0.47\textwidth]{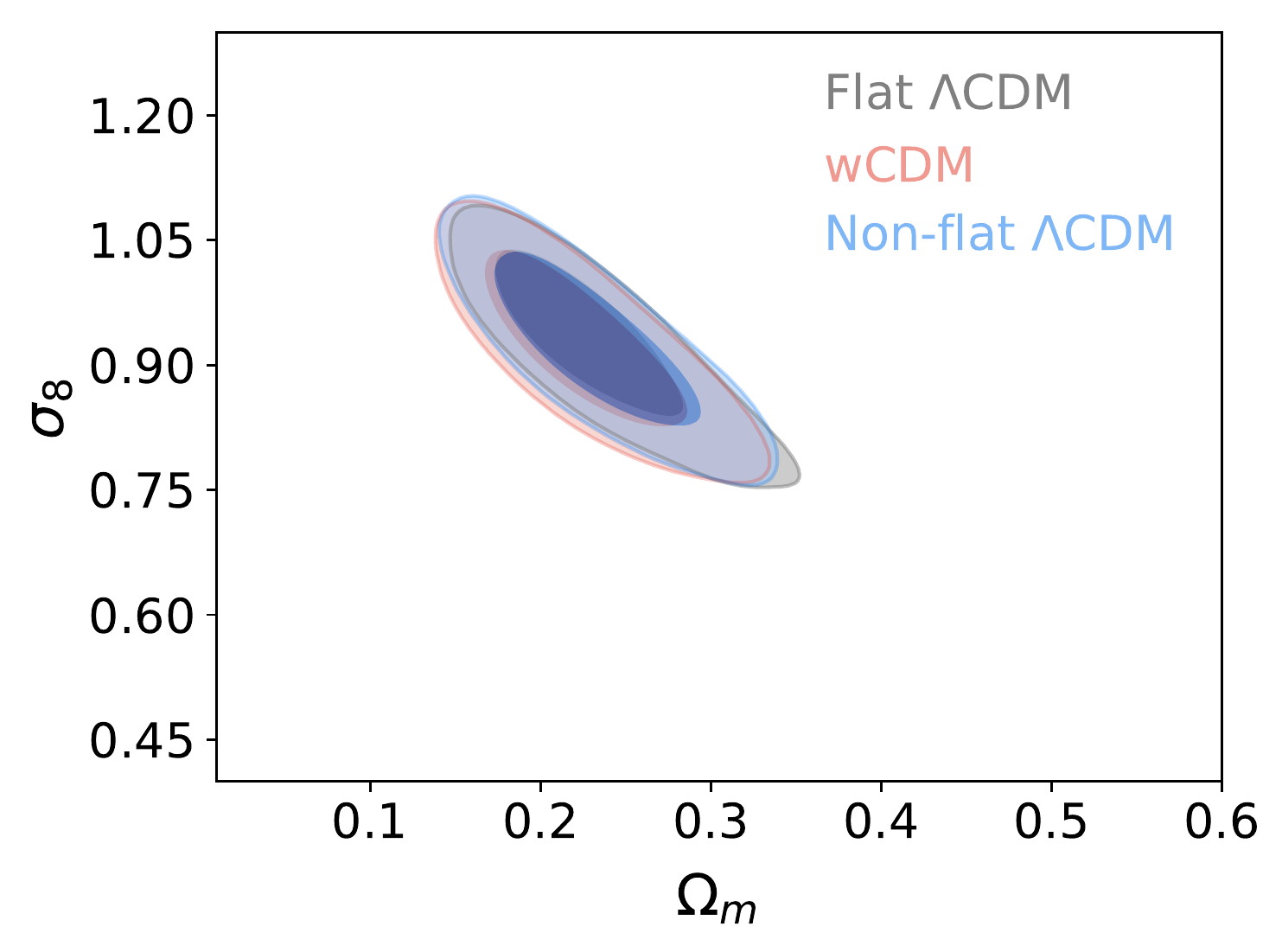}
    \caption{Constraints on $\sigma_8$ and $\Omega_m$ for the flat $\Lambda$CDM (grey), wCDM (red), and non-flat $\Lambda$CDM (blue) models. Little difference in parameter constraints is found among the three cosmological models.}
    \label{fig:one_param_ext_contour}
\end{figure}

Beyond our baseline cosmology, we
considered two one-parameter extension models to the flat $\Lambda$CDM cosmology, namely the non-flat $\Lambda$CDM and flat $w$CDM models.
For the non-flat $\Lambda$CDM model, we let $\Omega_k$ vary within the interval $[-0.2,0.2]$. We use a flat prior for the equation of state parameter $-1.5< w < -0.5$ when $w$CDM is assumed. Of course, weak lensing alone does not constrain these two parameters. The goal of this experiment is to investigate how the $\Omega_m$-$\sigma_8$ constraint changes as we assume these cosmologies.
We display the results in Figure~\ref{fig:one_param_ext_contour}, which shows that the variation among the three models is only a few tens of per cent of the statistical errors. The marginalized $\Omega_m$ and $\sigma_8$ values are summarized in Table~4.

\begin{deluxetable}{cccc}[h]
\tabletypesize{\scriptsize}
\tablecaption{Cosmological Parameter Constraints for Various Models}
\tablenum{4}
\tablehead{\colhead{model} & \colhead{$\Omega_m$} & \colhead{$\sigma_8$}& \colhead{$S_8$} }
\tablewidth{0pt}
\startdata
Flat $\Lambda$CDM & $0.231^{+0.021}_{-0.044}$ & $0.931^{+0.072}_{-0.054}$ & $0.810^{+0.039}_{-0.031}$\\
 & & & \\[-1.5pt]
Flat $\Lambda$CDM w/o baryon & $0.178^{+0.018}_{-0.029}$ & $0.984^{+0.061}_{-0.051}$ & $0.753^{+0.040}_{-0.030}$\\
 & & & \\[-1.5pt]
$w$CDM & $0.226^{+0.036}_{-0.039}$ & $0.925^{+0.064}_{-0.062}$ & $0.796^{+0.039}_{-0.044}$\\
& & & \\[-1.5pt]
Non-flat $\Lambda$CDM & $0.233^{+0.029}_{-0.046}$ & $0.925^{+0.064}_{-0.062}$ & $0.807^{+0.042}_{-0.042}$
\enddata
\tablecomments{The selected models are a flat $\Lambda$CDM with/without baryonic feedback, a $w$CDM ($-1.5 < w < 1.5$), and a non-flat $\Lambda$CDM ($ -0.2 < \Omega_k < 0.2$). We include AGN feedback and neutrino masses for both $w$CDM and non-flat $\Lambda$CDM.}
\label{tab:one_param_ext}
\end{deluxetable}

\section{Discussion} \label{sec:discussion}

\begin{figure*}
    \centering
    \includegraphics[width = 0.9\textwidth]{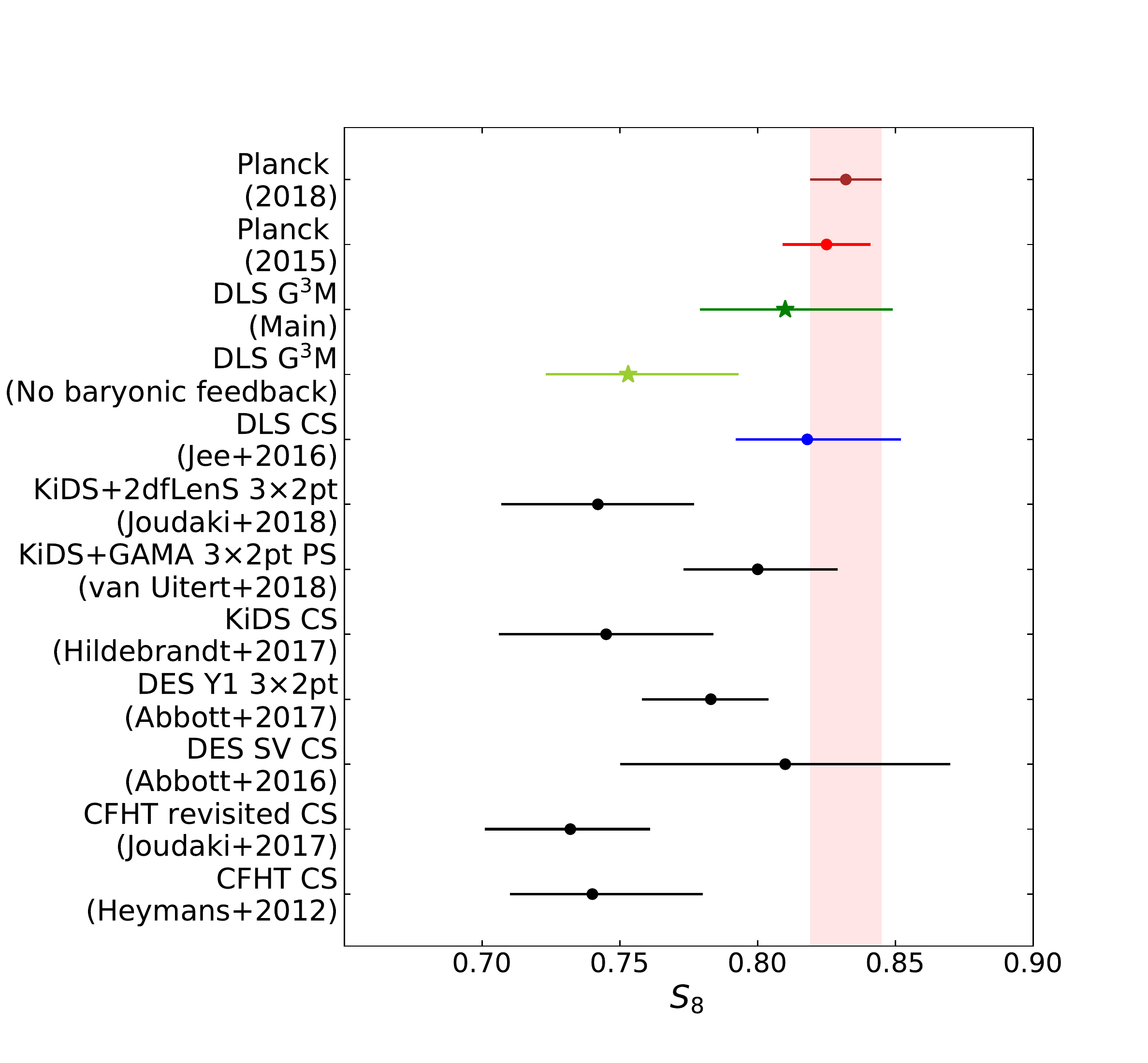}
    \caption{Comparison of $S_8$ values among different surveys: Planck2018 (brown), Planck2015 (red), DLS G$^3$M with $A_{baryon}$ marginalization (green), DLS G$^3$M without $A_{baryon}$ marginalization (light green),  DLS cosmic shear (blue), KiDS,  DES, and CFHT.}
    \label{fig:S8_in_surveys}
\end{figure*} 

\subsection{Comparison of the DLS $S_8$ Measurement with Other Studies}

Because of the well-known $\Omega_m$-$\sigma_8$ degeneracy, the $S_8\equiv \sigma_8 (\Omega_m/0.3)^{0.5}$ value is a popular choice for comparing results from different surveys. In this paper, we also use this $S_8$ parameter to enable comparison with previous studies. However, it is important to remember that the comparison is fair only to the extent that the measurement (i.e., the exact shape of the degeneracy) favors this particular choice of the exponent $0.5$.

The weak-lensing surveys that we use for comparison are the CFHTLenS \citep{2012MNRAS.427..146H,2017MNRAS.465.2033J}, DES Science Verification \citep{Abbott:2015swa}, DES Year 1 \citep{Abbott:2017wau}, and KiDS \citep{2017MNRAS.465.1454H, doi:10.1093/mnras/sty551}. The results from these surveys are compared with the DLS results in Figure~\ref{fig:S8_in_surveys}. Also displayed in Figure~\ref{fig:S8_in_surveys} are the Planck constraints \citep[][Planck temperature + low $\ell$ polarizations + lensing]{Ade:2015xua,Aghanim:2018eyx}.
The discrepancies between the $S_8$ values from some weak-lensing studies and the Planck CMB study and between the $H_0$ values from the direct measurements and the Planck CMB-inferred value (not shown)
are often referred to as a ``low-$z$ vs. high-$z$ tension". If the tension is real, one may interpret the difference as indicating a need for some extensions of the standard $\Lambda$CDM model and/or revision of astrophysical models. 
For example, \cite{2015MNRAS.451.2877M} made an attempt to explain the tension between CFHTLenS and Planck with several additional parameters such as intrinsic alignment, AGN feedback, neutrino mass and neutrino species, etc. However, they found that none of the efforts could relieve the tension significantly. \cite{doi:10.1093/mnras/stx998} showed the $w$CDM model is moderately preferred to relieve the tension in $S_8$, while $w$CDM relieves the tension in Hubble constant to some extent. 

Among the results shown in Figure~\ref{fig:S8_in_surveys}, the studies having a $\mytilde2\sigma$ tension with the Planck results are the CFHTLenS \citep{2012MNRAS.427..146H, 2017MNRAS.465.2033J}, KiDS+2dfLenS 3$\times$2pt \citep{Joudaki:2017zdt}, KiDS cosmic shear \citep{2017MNRAS.465.1454H} and DES Year 1 3$\times$2pt \citep{Abbott:2017wau} whereas the DES SV cosmic shear \citep{Abbott:2015swa}, KiDS+GAMA 3$\times$2pt power spectrum \citep{doi:10.1093/mnras/sty551}, and DLS studies do not show such a tension. The fact that some weak-lensing studies do not present any tension with the Planck CMB result may hint at the possibility that some surveys might have suffered from unknown systematics. For example, the two studies from KiDS produce somewhat different $S_8$ values. The KiDS tomographic cosmic shear analysis \citep{2017MNRAS.465.1454H} leads to $S_8=0.745\pm0.039$ whereas the KiDS study combining both cosmic shear and G$^3$M measurements  \citep{doi:10.1093/mnras/sty551} gives $S_8=0.800_{-0.027}^{+0.029}$. The statistical inconsistencies in KiDS are discussed in \cite{doi:10.1093/mnras/sty099}, who claim that it is too early to regard the tension as statistically meaningful.

\subsection{Impacts of Assumed Cosmology on Parameter Estimation}
\label{sec:cosmology_dependence}

In a few steps of our analysis, it is necessary for us to assume particular cosmological parameter values. 
They are the covariance matrix estimation (\S\ref{sec:cov_estimation}), the integral constraint computation in the galaxy auto-correlation measurement (\S\ref{sec:galaxy_angular_correlation}), and the ``tail" correction in correlation function evaluations (\S\ref{sec:ps_reconstruction}). Here we discuss the influence of the assumed cosmology.

The covariance matrix discussed in \S\ref{sec:cov_estimation} consists of four parts: the shot noise, systematic error, mixed term, and sample variance. Because the sample variance is a function of cosmology, in principle the likelihood evaluation (Equation~\ref{eqn:likelihood}) needs to compute the covariance matrix in each sampling.
In our study, we use the mock galaxy and shear catalogs from the FLASK package, whose resulting statistics follow log-normal distributions. Although this method is faster than the one that relies on $N$-body simulation data, it is still not feasible to implement the cosmology-dependent covariance.
The cosmology-sensitivity in parameter estimation has been discussed in many studies \citep[e.g.,][]{2009A&A...502..721E,2013MNRAS.430.2200K, Jee2013, 2013PhRvD..88f3537D}, and the results are somewhat inconclusive. Perhaps, the issues somewhat depend on the method for covariance generation and the characteristics of the survey data. 

We assess the impact of the cosmology dependence of the covariance matrix on our parameter estimation by repeating the analysis procedure using several covariance matrices generated at different cosmological parameters. From this limited test, we find that the results are mostly sensitive to the input $\sigma_8$ values. The central values do not differ much, showing no apparent correlation with the input $\sigma_8$ value, whereas the parameter errors certainly increase with $\sigma_8$. For example, in two extreme cases, where we set the input $\sigma_8$ to 0.6 and 1.05, the difference in the central value of $S_8$ is $\mytilde0.006$ whereas its error increases by $\mytilde30\%$. Since our best-fit values are in good agreement with the input cosmology, we believe that the amount of bias in the parameter estimation and their errors (from not including cosmology-dependent covariance) is negligible.

With a similar method, we test the impact of the selected cosmology in the ``tail" creation and IC evaluation. Since both the presence of the tail and the upshifting of the galaxy auto-correlation with IC values add to the amplitude of the power spectra at low $\ell$'s, we expect the central values of the estimated parameters to correlate with the input $\sigma_8$ value to some extent. Indeed, we find such a tendency in our experiment, although the difference is still smaller than the statistical errors. For example, the use of the input values $\sigma_8=0.6$ or 1.05 leads to a $\mytilde0.004$ shift in $S_8$ ($\mytilde12$\% of the statistical error) compared to the result when the input value is $\sigma_8=0.83$. 

\subsection{Constraints on Baryonic Feedback Parameter and Model Selection}
\label{sec:baryon_model_sel}

\begin{figure}
    \centering
    \includegraphics[width = 8.5cm]{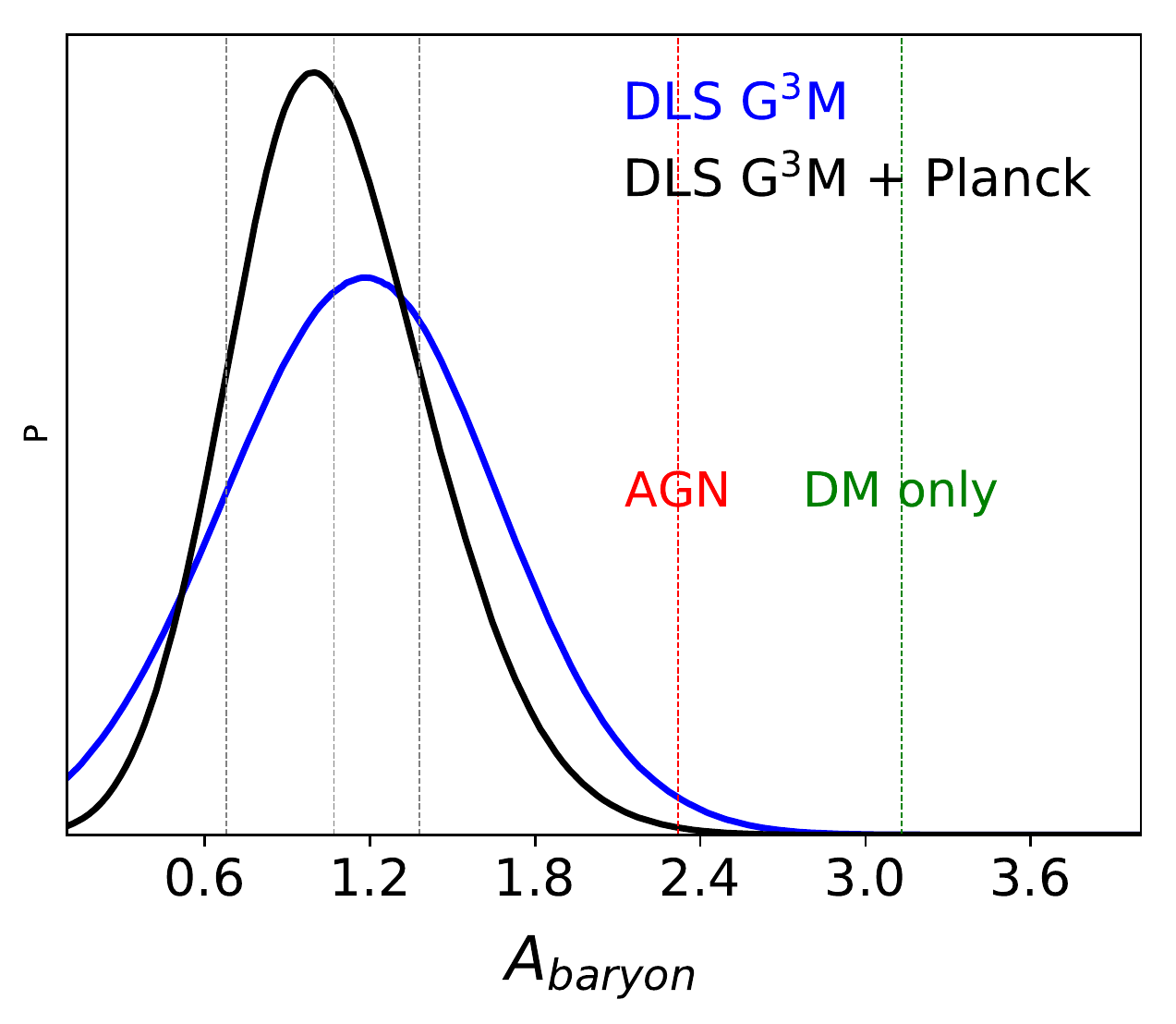}
    \caption{Constraint on the baryonic feedback parameter using the DLS alone (blue) and when combined with Planck2015 data (black). The solid curves are obtained after we apply a Gaussian kernel to smooth the marginalized one-dimensional histogram of $A_{baryon}$. The left, middle, and right gray dashed lines represent the 1-$\sigma$ lower limit, the mean, and the 1-$\sigma$ upper limit, respectively.
    The red and green lines show the $A_{baryon}$ values from Mead et al. (2015) corresponding to the AGN feedback and dark matter-only cases, respectively.}
    \label{fig:A_baryon}
\end{figure}

As shown in \S\ref{sec:param_result} and Figure~\ref{fig:S8_in_surveys}, our introduction of the baryonic feedback parameter $A_{baryon}$
leads to a higher value of $S_8$ than without it. This behavior is easy to understand because the net effect of the AGN feedback lowers the amplitude of the power spectrum. 
Within the prior interval $2<A_{baryon}<4$ we cannot constrain the parameter value. The posterior probability distribution shows that
only the upper limit of $A_{baryon}$ is bound within the interval. 
The probability gradually increases as $A_{baryon}$ approaches the lower limit $A_{baryon}=2$, which indicates that we might need to 
extend the prior interval in order to constrain the lower limit of $A_{baryon}$ as well. Here, we present the result from our experiment on the $A_{baryon}$ measurement using an extended prior interval $[0.1,4]$. Since the $A_{baryon}\lesssim2$ regime has not been validated with corresponding numerical simulations, we must use caution when interpreting the results.

Using the DLS G$^3$M data alone, we are able to measure $A_{baryon}=1.19^{+0.51}_{-0.45}$ (Figure~\ref{fig:A_baryon}).
The resulting $S_8$ value increases by $\mytilde 1~\sigma$ (Appendix F), still consistent with the measurement of Planck2015. 
Since $A_{baryon}$ is degenerate with other cosmological parameters, an improved constraint is possible with the addition of external data.
We choose the Planck CMB data (Planck2015) because it provides independent, tight constraints on a number of cosmological parameters including $\Omega_m$ and $\sigma_8$, which significantly increases the power to constrain $A_{baryon}$ among other probes. We use the publicly available Planck likelihood code (Plik lite temperature + polarization)\footnote{\url{http://pla.esac.esa.int/pla/##cosmology}}. As a result, we obtain $A_{baryon}=1.07^{+0.31}_{-0.39}$ (Figure~\ref{fig:A_baryon}), which is highly consistent with the result when the DLS G$^3$M measurement is used alone. 

The constrained values are significantly smaller than the $A_{baryon}=2.32$ value that corresponds to the baryonic effects with the AGN feedback in \cite{2015MNRAS.454.1958M}. Note that this result is different from the conclusion of \cite{2015MNRAS.451.2877M}. Their analysis with the combination of the CFHTLens cosmic shear signal with the Planck CMB data does not show any preference for the power spectrum with the AGN feedback.
Our result may hint at the possibility that actual AGN feedback might be stronger than the OWLS AGN feedback prescription. However, we caution that this interpretation is limited by the validity of this one-parameter representation (Eqn.~\ref{eqn_baryon}) of the baryonic feedback effect for the power spectrum evaluation. As shown by \cite{2018arXiv180108559C}, both the amount of suppression and the scale where the effect is most significant vary among different cosmological hydro-simulations. When we consider the three state-of-the-art simulations: Horizon-AGN, OWLS, and Illustris, the suppression from Illustris is most severe with the maximum $\mytilde35$\% reduction with respect to the dark matter-only power spectrum at $k\sim6~h~\mbox{Mpc}^{-1}$. The maximum amount of suppression in OWLS is $\mytilde30$\% at $k\sim10~h~\mbox{Mpc}^{-1}$. For the Horizon-AGN case, although the exact angular scale where the maximum suppression occurs is similar to that of OWLS, the amount of suppression is less than $\mytilde15$\%. 

The amount of the power spectrum suppression for our $A_{baryon}\sim1$ case cannot be compared to the Illustris power spectrum
directly because the difference is a sensitive function of $k$.
At $k\sim1~h~ \mbox{Mpc}^{-1}$, we find that the amount of suppression corresponding to $A_{baryon}\sim1$ is similar (about 80\% suppression with respect to the DM-only case) to that of Illustris. On a larger scale the HMcode suppression with $A_{baryon}\sim1$ becomes weaker while on a smaller scale the trend is reversed. However, for the scales relevant for the G3M power spectra, the integrated suppression would be stronger for Illustris because its suppression starts to occur at smaller $k$ values. 

It is possible that the baryonic feedback parameter may trade off with other cosmological parameters. Any strong degeneracy between parameters can lead to incorrect
interpretation.
For example, \cite{Harnois-Deraps:2014sva} claim that the effect of neutrino mass is degenerate with baryonic feedback and thus cannot be ignored in cosmological parameter estimation. Currently, the two effects are measured separately from independent simulations. In this paper we implement the combined influence by multiplying the two effects without explicitly accounting for their possible degeneracies and covariances. Nevertheless, numerical studies show that the correlation between baryonic feedback and neutrino free streaming is negligible \citep[e.g.,][]{jing2006,daalen2011,bird2012}. Also we demonstrate in Appendix B, the amount of the power spectrum suppression due to neutrino is subdominant compared to the baryonic feedback effect. Therefore, the impact of neutrino is insignificant in our $A_{baryon}$ measurement.

Together with the above $A_{baryon}$ value constraint, another useful exercise is to test whether or not we can differentiate models with and without AGN feedback using the following Bayes factor:
\begin{equation}
BF=\frac{P(M_1|\mathbf{D})}{P(M_2|\mathbf{D})}     
\end{equation}
\noindent
where $P(M_1|\mathbf{D})$ and $P(M_2|\mathbf{D})$ are the probabilities of the $M_1$ and $M_2$ models given data $\mathbf{D}$. Since $P(M|\mathbf{D})=P(\mathbf{D}|M) P(M)$, 
evaluation of the above BF is performed using the Bayesian evidence $P(\mathbf{D}|M)$ with the assumption $P(M_1)=P(M_2)$. The computation of the evidence involves integrals of the likelihood in the parameter space $\boldsymbol{\theta}$ over wide intervals:
\begin{equation}
P(\mathbf{D}|M)=\int P(\mathbf{D}|M,\boldsymbol{\theta}) P(\boldsymbol{\theta}|M) d \boldsymbol{\theta},   
\end{equation}
\noindent
which is computationally more challenging than parameter estimation. We use the \texttt{multinest} package mentioned in \S\ref{sec:likelihood} to carry out this integration. The evidence value also depends on whether or not we  marginalize
over neutrino masses and we measure them separately.
Using the DLS G$^3$M data with (without) marginalizing over neutrino masses, we find the difference in the log evidences of the two models (dark matter-only vs. AGN feedback) to be $\mytilde2.0$ ($\mytilde3.1$), which implies that the model with the inclusion of the baryonic effects with AGN feedback is preferred at a moderate level. This result is in slight contrast with the study of \cite{2017MNRAS.465.2033J}, who claim from the re-analysis of the CFHTLenS data that their cosmological parameter constraints do not show any preference between the two models. 
When we combine the current DLS data with the Planck CMB constraint, the difference in the log evidence becomes $\mytilde10.7$, strongly favoring the power spectrum with AGN feedback; in this latter case the evidence difference estimate is not affected by the inclusion of the neutrino mass marginalization.

\section{Summary and Conclusions} \label{sec:conclusion}

We present cosmological parameter constraints by measuring galaxy-galaxy and galaxy-mass power spectra from the DLS. The power spectra are constructed using two lens bins at $\left< z \right >=0.27$ and 0.54 and two source bins at $\left< z \right >=0.64$ and 1.09 for the multipole range $\ell=250\sim2000$. Our lens-source flip and B-mode tests do not reveal any significant systematic errors in photo-$z$ and shear estimation. We address potential residual photo-$z$ and shear calibration systematics by marginalizing over one shear calibration and four photo-$z$ bias parameters in our cosmological parameter constraint. Also, we account for the power spectrum suppression due to both AGN feedback and neutrinos by employing the power spectrum model that includes the effects and marginalizing over the feedback and neutrino mass parameters.

The $S_8=\sigma_8 (\Omega_m/0.3)^{0.5}$ value is constrained to $S_8 = 0.810^{+0.039}_{-0.031}$. This value is in excellent agreement with our previous estimate from the DLS cosmic shear study. We expect that the cosmic shear-based $S_8$ value would increase somewhat when the baryonic feedback effect is included. Although the uncertainty of $S_8$ in the current study is slightly ($\mytilde20$\%) larger than cosmic shear result, the $\Omega_m$-$\sigma_8$ degeneracy is reduced by $\mytilde40$\%. Our result does not cause any tension with the value derived from the latest Planck measurement. 

Our galaxy bias values are also well-constrained and show marginal evidence for redshift evolution; galaxies at higher redshift have larger bias. We examine the internal consistency by independently determining biases using the fixed, best-fit cosmology. The test shows that the results from both galaxy-galaxy and galaxy-mass power spectra are consistent with each other, although the signal for the possible redshift-evolution mostly comes from the galaxy-galaxy power spectrum.

The Bayesian evidence with the DLS-only case indicates that the power spectrum model with baryonic feedback is preferred at the moderate level. The combination of the DLS data with the Planck CMB measurements strongly favors the power spectrum with AGN feedback.
We find that the best-fit $S_8$ value decreases by $\mytilde0.05$ when we use a dark matter-only power spectrum. Considering the size of the parameter uncertainty ($\mytilde0.04$) and the angular scale of the power spectrum suppression, we believe that the difference is non-negligible.

Combining the current galaxy-galaxy and galaxy-mass power spectra with the Planck CMB data, we are able to constrain the baryonic feedback parameter to $A_{baryon}=1.07^{+0.31}_{-0.39}$. This value is significantly smaller than the fiducial value $A_{baryon}=2.32$, which is derived by matching the revised halo model power spectrum to the OWLS results with AGN feedback. Our $A_{baryon}$ constraint may hint at the possibility that the recipe used in the OWLS simulation might have been weaker than actual AGN feedback. However, the interpretation is tentative until we verify the validity of this one-parameter representation of the baryonic feedback effect.

\acknowledgments
{We thank the anonymous referee for his/her constructive suggestions for improving the quality of the paper.
We also thank Kyle Finner and Cris Sabiu for their careful reading of the manuscript and providing useful comments.
M. Yoon acknowledges support from the Yonsei University Observatory -- KASI Joint Research Program (2018). M.J. Jee acknowledges support for the current research from the National Research Foundation of Korea under the program 2017R1A2B2004644 and 2017R1A4A1015178.}

\bibliographystyle{apj.bst}
\bibliography{dls_ggl_cosmology} 

\appendix

\section{Appendix A. Power spectrum comparison with/without Baryonic feedback }
\label{sec:appendixc}
\begin{figure}
\centering
    \includegraphics[width = 0.492\textwidth]{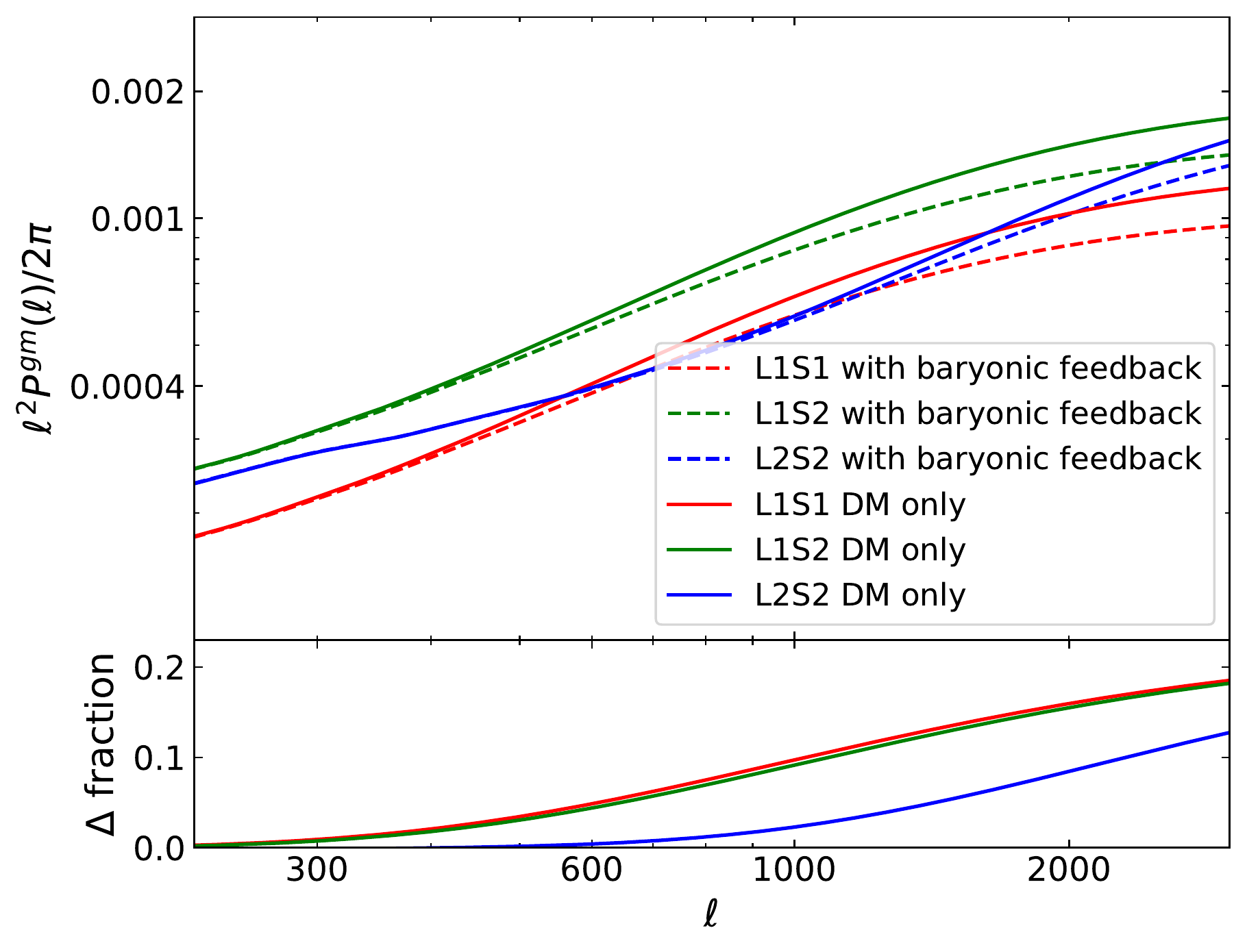}
    \includegraphics[width = 0.475\textwidth]{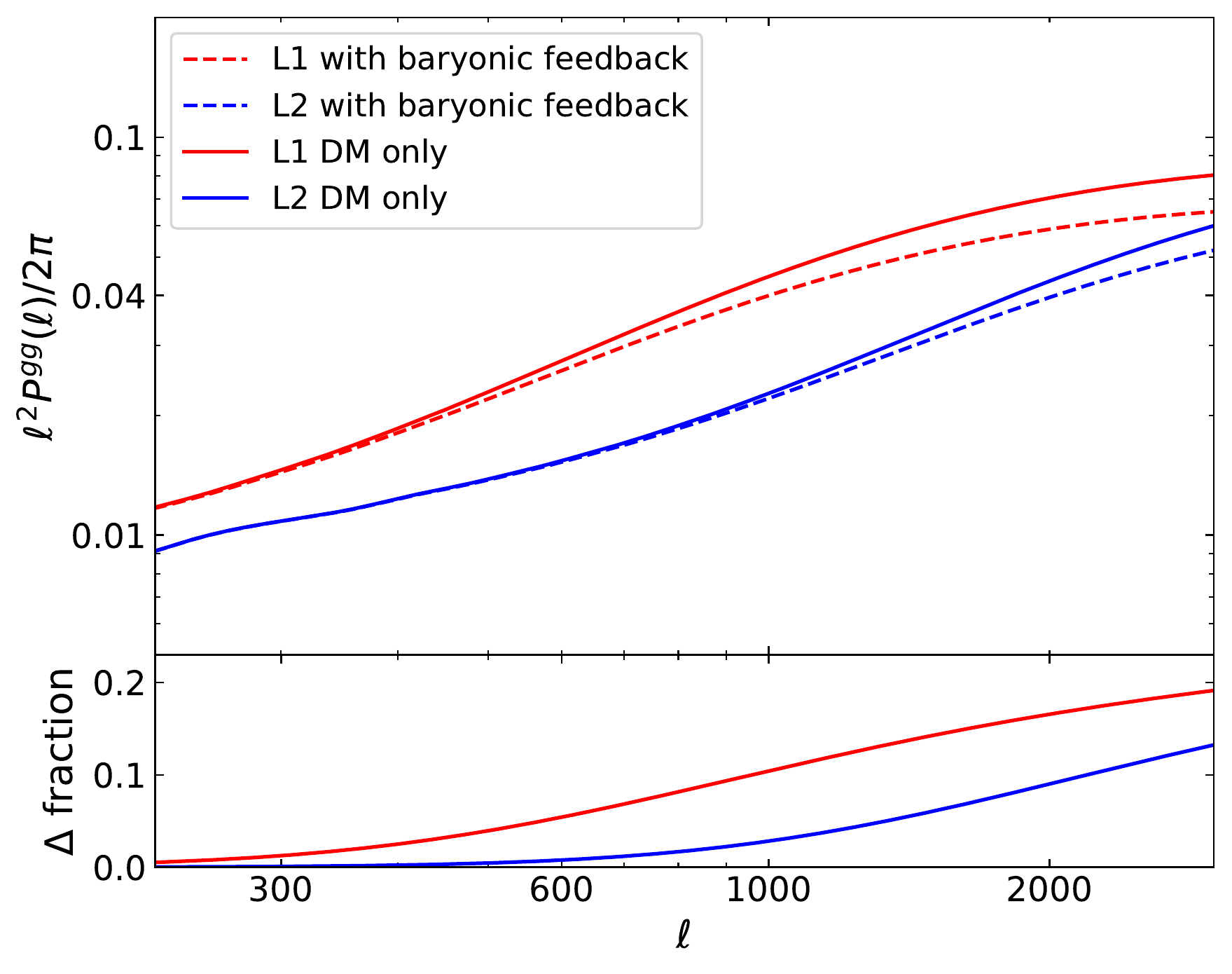}
    \caption{Comparison of galaxy-galaxy ($P^{gg}$) and galaxy-mass ($P^{gm}$) power spectra with (dashed) and without (solid) AGN feedback. The bottom panel shows the fractional difference between the two with respect to the dark matter-only model. We fix the cosmological parameters to our best-fit results.  Following \cite{2015MNRAS.454.1958M}, we use $A_{baryon}=2.32$ and 3.13 for the cases with and without AGN feedback, respectively. The suppression of the power at large $\ell$ values is significant, reaching up to $\sim$18\% at $\ell\sim2000$.}
\label{fig:baryon_effect_inps}
\end{figure}

A common method to deal with unknown baryonic effects on the model power spectrum has been removal of signals on small scales, which results in significant loss of the survey S/N. In this paper, we choose to address the issue by using the \cite{2015MNRAS.454.1958M} power spectrum to control the degree of baryonic feedback using the single parameter $A_{baryon}$. In Figure~\ref{fig:baryon_effect_inps}, we show the $P^{gg}$ and $P^{gm}$ power spectrum shifts due to the baryonic effects including AGN feedback. 
We use $A_{baryon}= 2.32$ to represent the case of the baryonic effects with AGN feedback, which is the best-fit result to the OWLS simulation according to \cite{2015MNRAS.454.1958M}; the dark matter-only case corresponds to $A_{baryon}= 3.13$. It is clear that the suppression of the power at large $\ell$'s is significant and up to $\sim$18\% at $\ell\sim2000$.

\section{Appendix B. Power spectrum comparison with/without massive neutrino }

\begin{figure}
\centering
    \includegraphics[width = 0.492\textwidth]{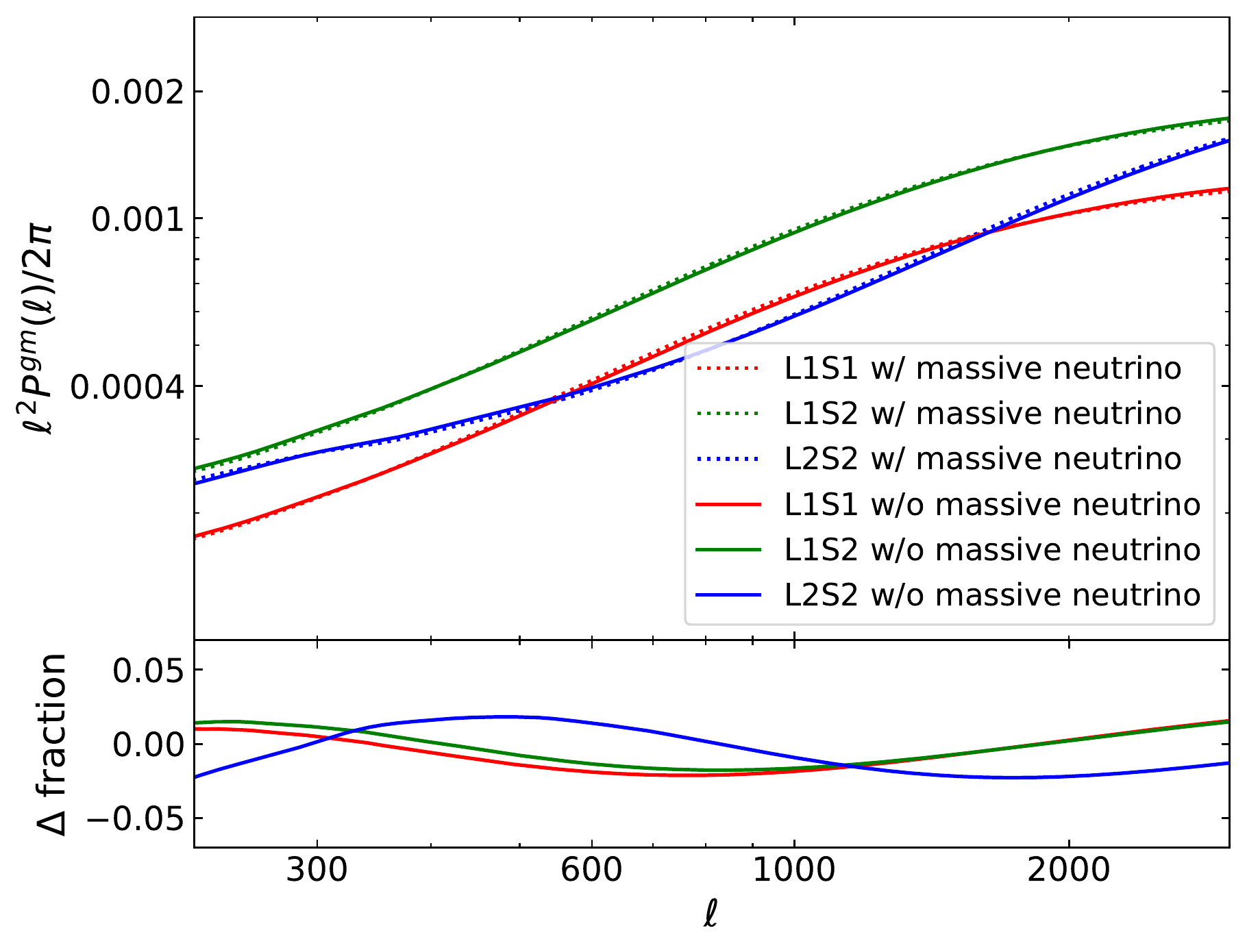}
    \includegraphics[width = 0.475\textwidth]{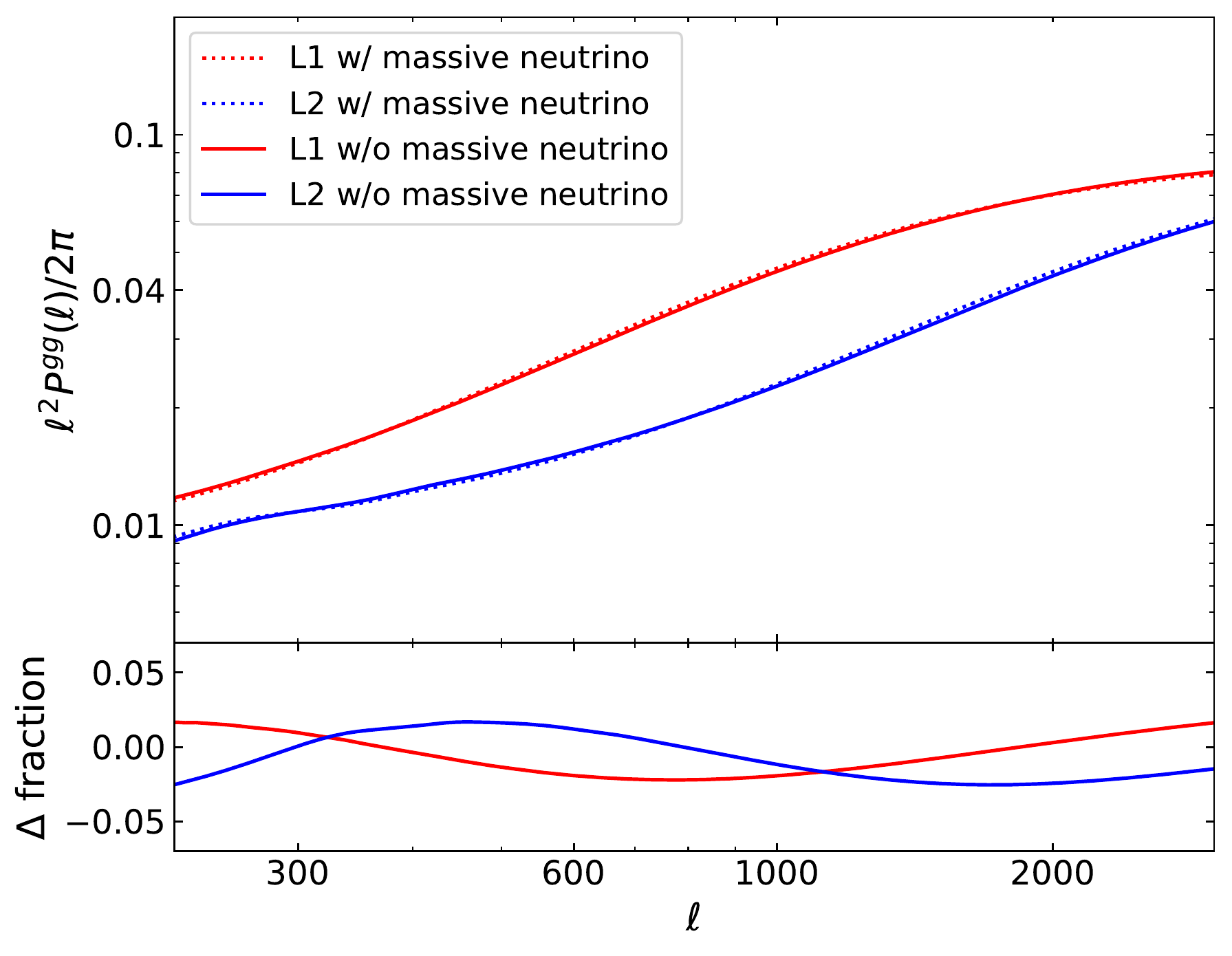}
    \caption{Comparison of galaxy-galaxy ($P^{gg}$) and galaxy-mass ($P^{gm}$) power spectra without (solid) and with (dotted) massive neutrinos ($\Sigma_{\nu} m_\nu  = 0.6~$eV). The bottom panel shows the fractional difference with respect to the case without massive neutrinos. We fix the cosmological parameters to our best-fit results.}
\label{fig:neutrino_effect_inps}
\end{figure}

Similarly to baryonic feedback, massive neutrinos also suppress the power on small scales. That is, the baryonic feedback effect is degenerate with the effect played by massive neutrinos. Here we illustrate how much our $P^{gg}$ and $P^{gm}$ power spectra are affected by massive neutrinos.
Figure \ref{fig:neutrino_effect_inps} shows the impact of  massive neutrinos on the galaxy-galaxy and galaxy-mass power spectra for the case $\Sigma_{\nu} m_{\nu}=0.6$~eV, which
approximately corresponds to the 95\% upper limit constrained by Planck2015.
The maximum departure from the dark matter-only case (without AGN feedback) is $\mytilde3$\%, given the same matter power spectrum normalization $\sigma_8$. Note that neutrinos in general suppress power on small scales. However, when we choose to normalize the power spectrum with neutrinos in such a way that the result gives the same $\sigma_8$ value from the case with zero neutrino mass, the resulting shift is both positive and negative depending on scales.\\

\section{Appendix C. L1 and L2 Redshift Distribution Calibration}

It is generally agreed that the photo-$z$ bias of a galaxy population is reduced when one constructs the population's $p(z)$ by stacking $p(z)$ of individual galaxies rather than point estimates \citep[e.g.,][]{2009ApJ...700L.174W}. 
A further reduction of the bias can be done through  comparison of photo-$z$ data with spectroscopic catalogs. For DLS, this photo-$z$ calibration is possible for L1 and L2. In terms of both magnitude and redshift ranges, the PRIMUS catalog is nearly complete for both L1 and L2 whereas the SHELS catalog is complete for L1. Thus, we use  only the PRIMUS catalog for L2 and both catalogs for L1.
For L1 we have 5,647 and 1,749 matching galaxies from SHELS and PRIMUS, respectively. On the other hand, we find 2,488 spectroscopic objects for L2. We note that this kind of the $p(z)$ calibration is not feasible for S1 and S2 because of the incompleteness of the spectroscopic catalogs.

\begin{figure}
\centering
    \includegraphics[width = 0.495\textwidth]{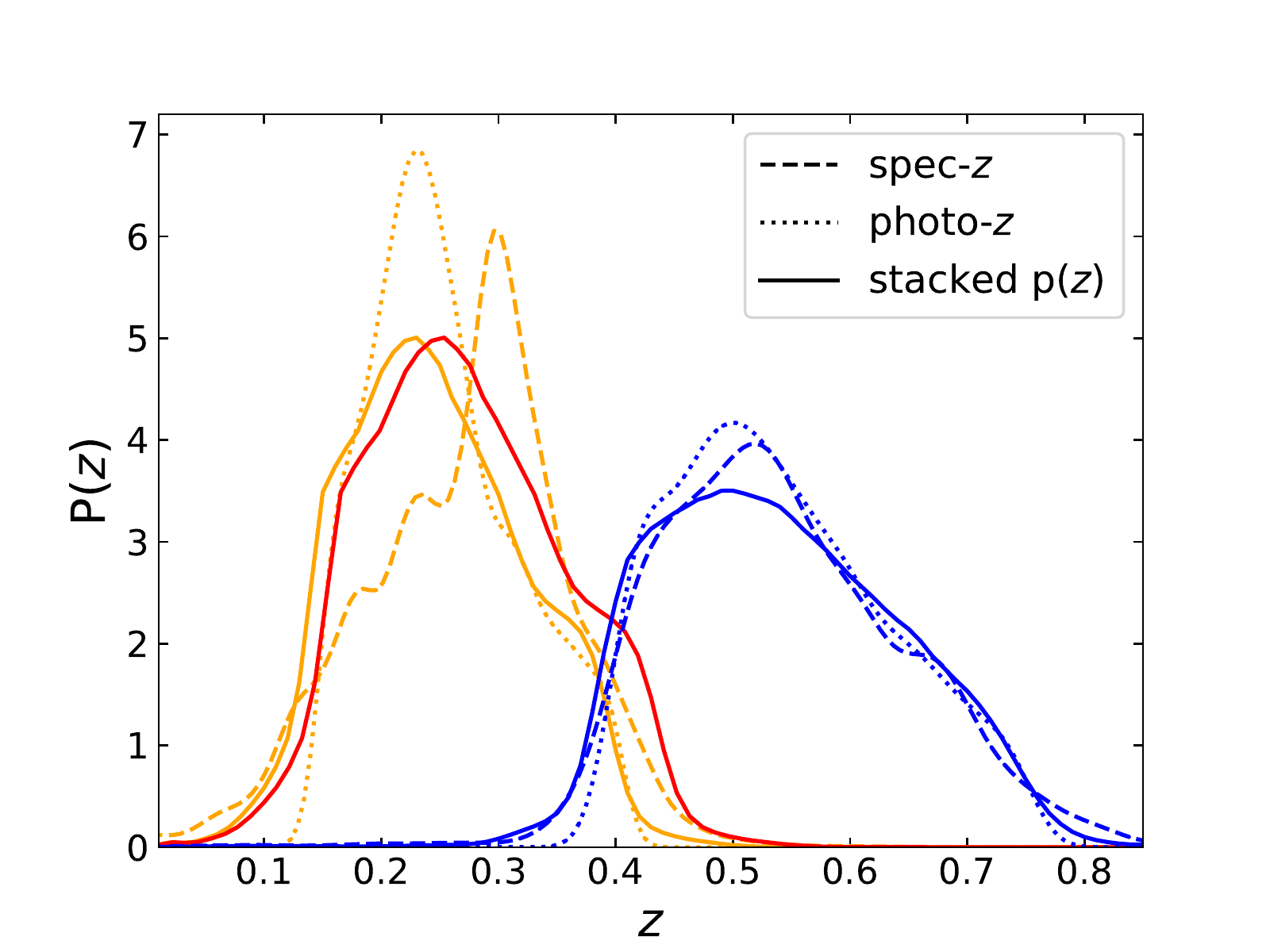}
    \caption{Calibration of the DLS photometric redshift distribution with spectroscopic data. We compare the population $p(z)$ distribution constructed from spec-$z$'s, point-estimate photo-$z$'s, and stacked $p(z)$ curves. We use orange and blue colors to represent the L1 and L2 populations, respectively. The bias is negligible for L2 whereas it is not for L1.  The mean redshift of the L1 population would be underestimated by $\mytilde10$\% if left uncorrected. The red solid curve shows our correction made by stretching the $p(z)$ curve horizontally so that the resulting mean agrees with the one from the spec-$z$ catalog. As noted in the text, this $p(z)$ calibration leads to the reduction of $S_8$ by $\mytilde0.02$, which is $\mytilde50$\% of the statistical error.
    }
\label{fig:photo_z_cal}
\end{figure}

Figure \ref{fig:photo_z_cal} compares the population $p(z)$ constructed from the point-estimate photo-$z$'s, spec-$z$'s, and stacked $p(z)$ of individual galaxies. We use the Kernel Density Estimator (KDE) to obtain the smooth $p(z)$ curves of the point-estimate photo-$z$'s and spec-$z$'s. We find that the bias is non-negligible for L1. The mean redshift of the L1 population would be underestimated by $\mytilde10$\% if left uncorrected whereas the agreement between photo-$z$ and spec-$z$ is excellent for the L2 galaxies (the difference in the mean is less than 1\%). The large discrepancy for the L1 population is caused by severe degeneracies for galaxies reported to be $z_b<0.4$ by BPZ. The lack of a $U$ filter in the DLS is known to be one of the main sources of the degeneracy in this redshift range. In this study, we address the issue by stretching the redshift range of the stacked $p(z)$ curve using Equation~\ref{eqn_photo_z} so that the resulting mean matches the spectroscopic value. We find that this $p(z)$ calibration results in the reduction of $S_8$ by $\mytilde0.02$ compared to the case without the calibration. The amount of the shift corresponds to $\mytilde50$\% of the statistical error.

Even after the above $p(z)$ calibration, the difference in the $p(z)$ shape still remains. Thus, we considered completely replacing the $p(z)$ with the spectroscopic $p(z)$ and found that our cosmological parameters virtually remain unchanged. Nevertheless, we think that this complete replacement lacks justification because the spectroscopic sample is only available to F2 and F5. 

\begin{figure}
\centering
    \includegraphics[width = 0.495\textwidth]{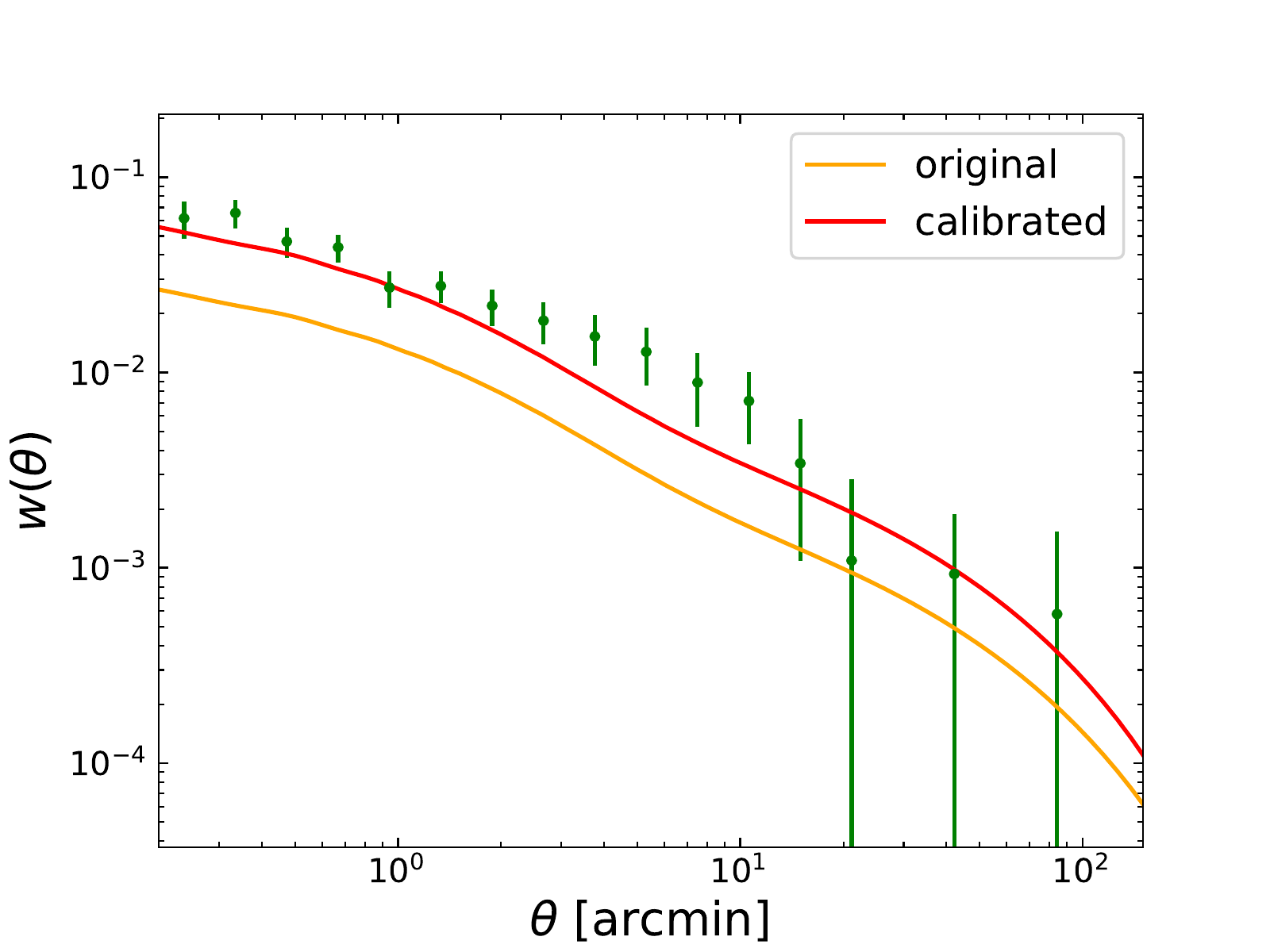}
    \caption{Galaxy-galaxy cross-correlation between L1 and L2. The green dots with error bars show the direct measurements. The amplitude should not be zero as shown because of the overlap in $p(z)$ between the two populations. The uncalibrated L1 $p(z)$ curve (orange solid line in Figure~\ref{fig:photo_z_cal}) does not sufficiently overlap with the $p(z)$ distribution of L2 and the predicted cross-correlation (orange) is significantly lower than the observation. Our $p(z)$ calibration (red) remarkably improves the agreement. When we predict the cross-correlation using the spec-$z$ $p(z)$ (orange dashed line in Figure~\ref{fig:photo_z_cal}), the values are virtually identical to the results from the calibrated $p(z)$.}
\label{fig:photo_z_cross}
\end{figure}

One powerful method to test the fidelity of this $p(z)$ calibration is to measure galaxy cross-correlation signals between L1 and L2 and compare them with the theoretical prediction based on these calibrated $p(z)$ curves. Figure~\ref{fig:photo_z_cross} shows the remarkable agreement between the theoretical cross-correlation function and the measurement. Also displayed is the prediction based on the uncalibrated $p(z)$ curve, which is clearly offset from the measurement. The increase in the predicted cross-correlation is due to the enlarged overlap in $p(z)$ between L1 and L2.
This cross-correlation test serves as a verification of our $p(z)$ calibration. We note that although one may consider using the cross-correlation measurements as additional constraints, in this study we only employ them for our $p(z)$ calibration verification.

\section{Appendix D. Impact of Random Signal Subtraction on Tangential Shear Measurement}

\begin{figure}
\centering
    \includegraphics[width = 0.47\textwidth, trim=0.5cm 0cm 0cm 0cm]{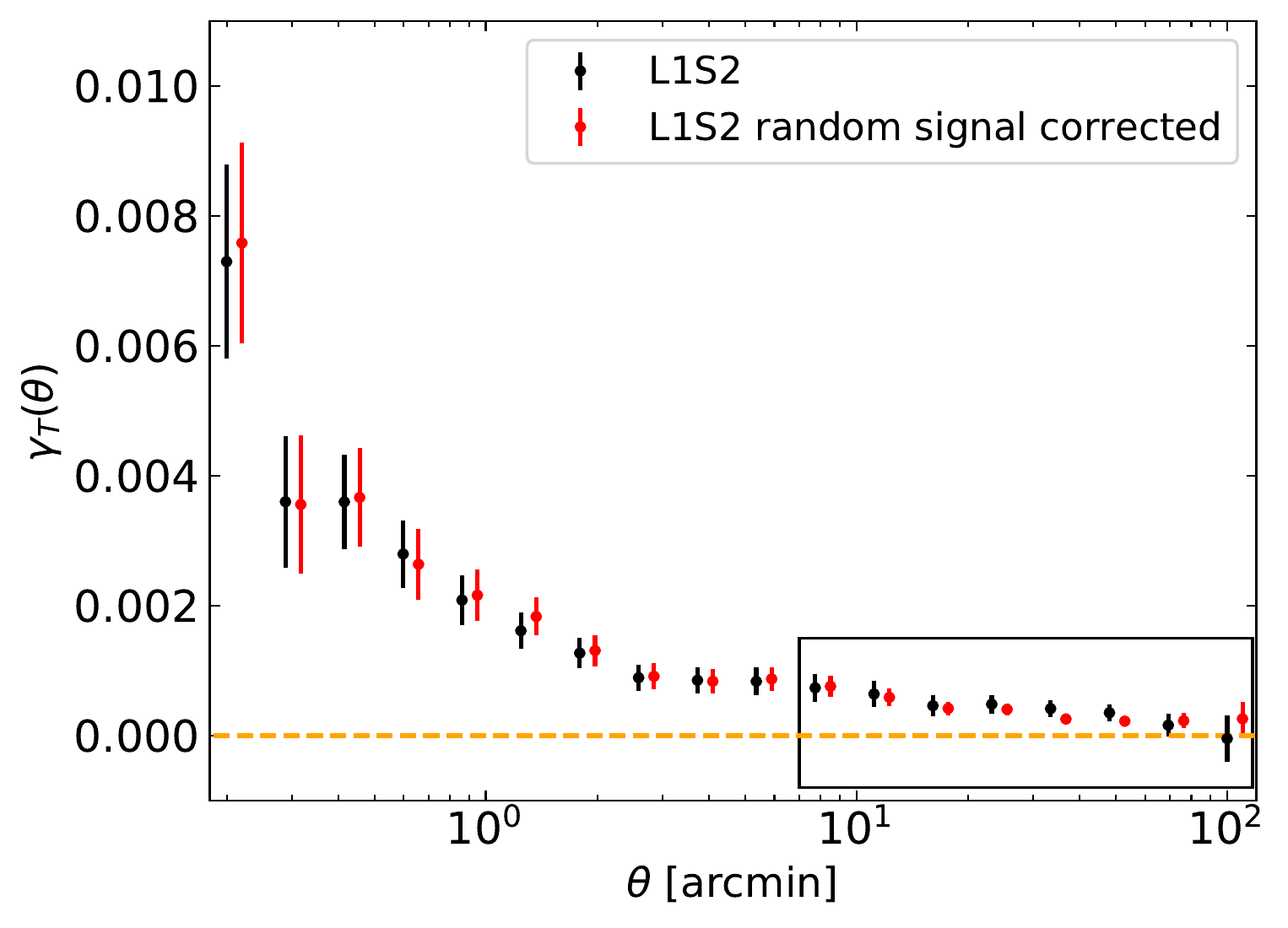}
    \includegraphics[width = 0.5\textwidth]{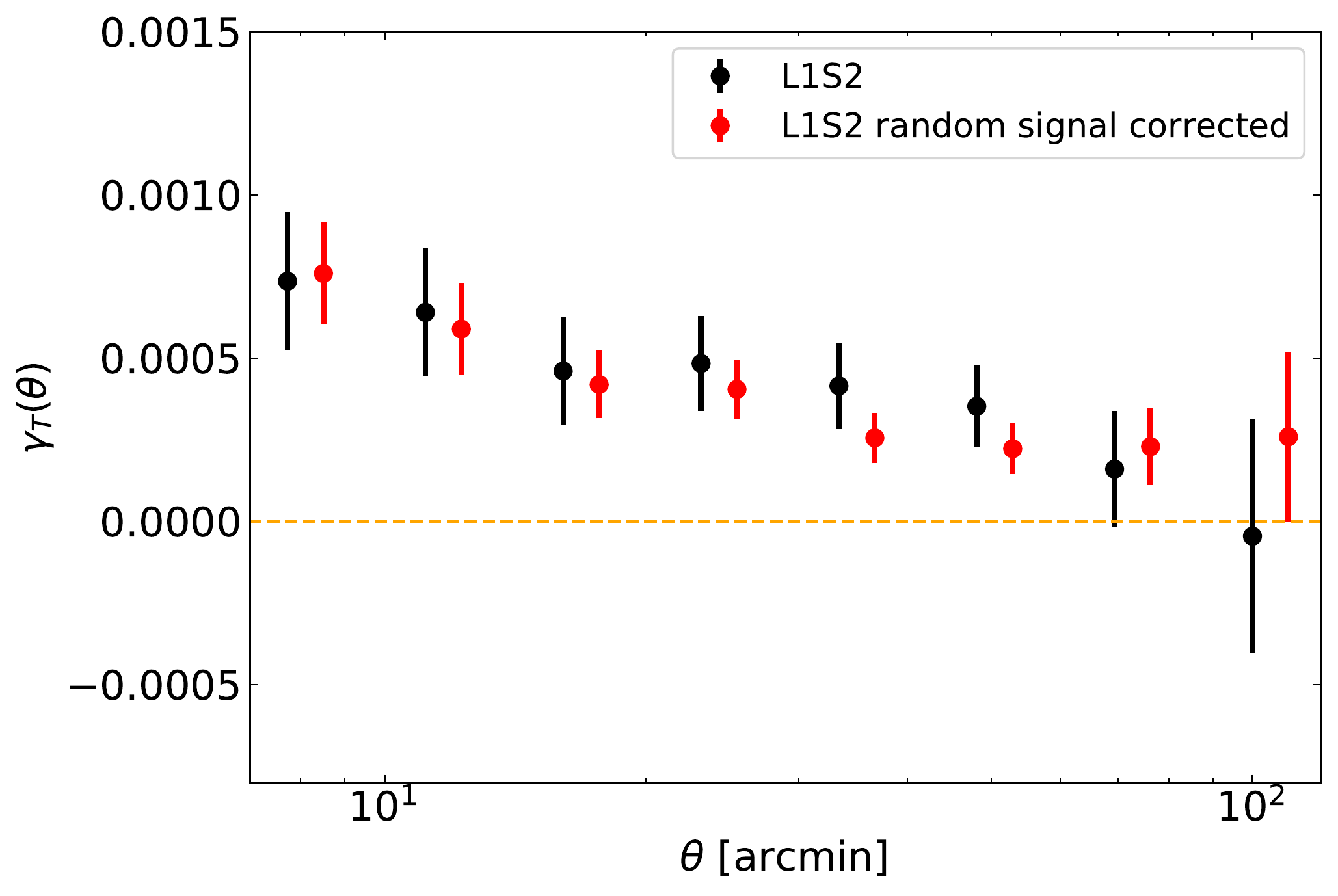}
    \caption{Tangential shear comparison between the cases with (red) and without (black) random signal subtraction. The displayed case is for the L1S2 pair. The right panel is a zoomed-in version showing the boxed region of the left panel. As shown, the random signal subtraction reduces both statistical and systematic errors caused by survey boundaries and star masking.}
\label{fig:random_subtraction}
\end{figure}

When tangential shears are measured, in principle, averaging over many lens-source galaxy pairs reduces/cancels additive shear biases. However, in practice, irregular sky coverage due to field boundaries and stellar masking regions hampers this bias reduction and increases field-to-field signal variations. The situation can be remedied by subtraction of tangential shears measured from randomly distributed (lens) points \citep{Singh:2016edo}. We show the effect of this random signal subtraction in Figure \ref{fig:random_subtraction}. With this correction, the size of the errors decreases and also the central values shift as shown. The change is more noticeable at large angles. The DLS is composed of five fields (F1-F5) and thus biases introduced by different observational footprints cause the tangential shear measurements of different fields to deviate considerably from one another, which increases statistical errors when they are averaged. After the correction, the tangential shear measurements from different fields become more consistent with one another. 

This correction is applied not only when we measure the tangential shear from the DLS but also when we estimate covariance from FLASK simulations. We apply the DLS star masking and the field boundaries to the simulated fields and subtract the signals measured from the pairs of random points and the simulated source galaxies. As tested in \cite{Singh:2016edo}, this provides a useful method to estimate an unbiased covariance taking account of field boundaries and star masking.

\section{Appendix E. Constraint on Intrinsic Alignment with the L2-S1 Power Spectrum}
\label{sec:appendixe}
\begin{figure}
\centering
    \includegraphics[width = 0.42\textwidth]{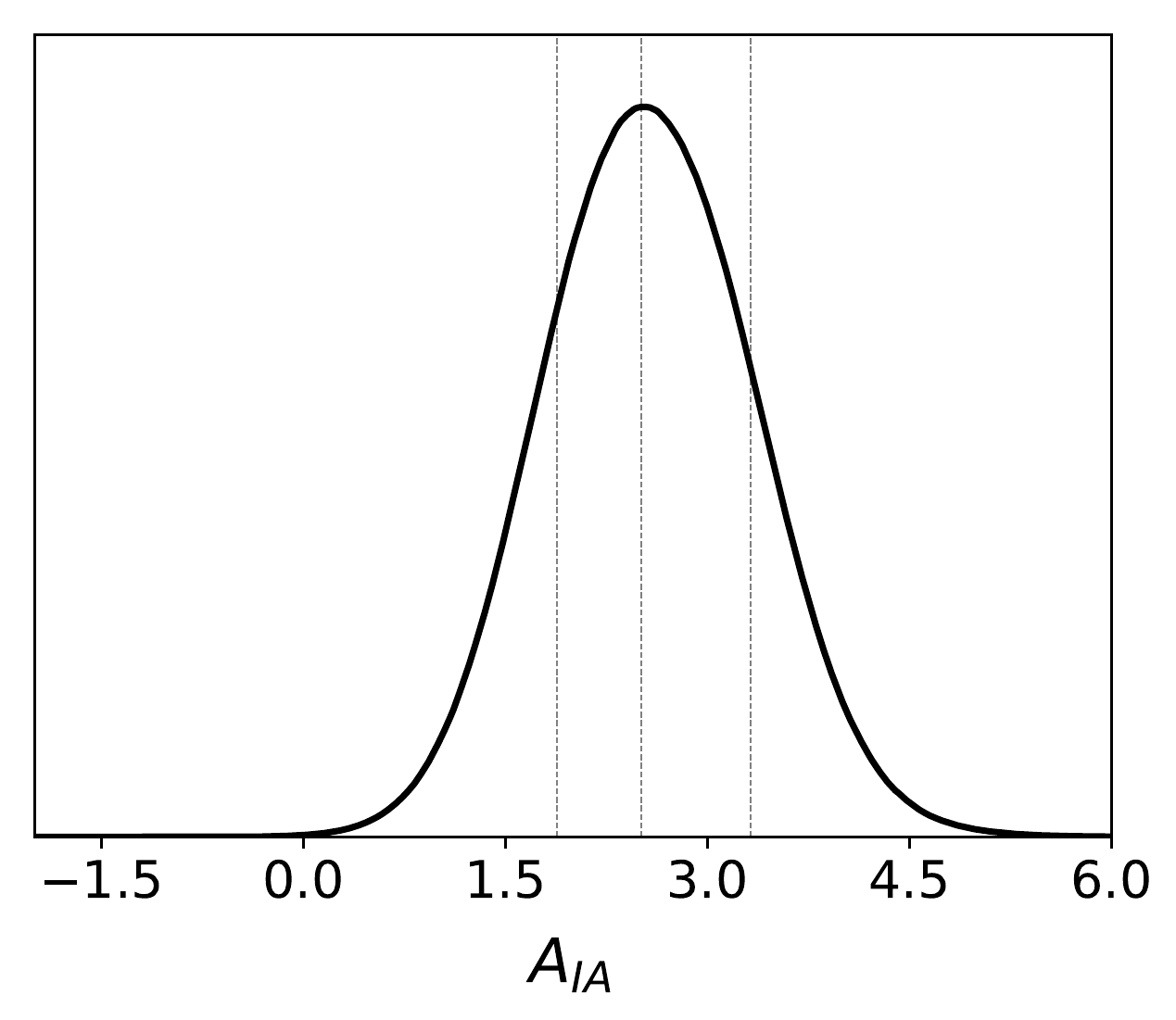}
    \caption{Constraint on intrinsic alignment amplitude. When we include the L2-S1 pair, we are able to constrain the intrinsic alignment amplitude $(A_{IA} = 2.51^{+0.82}_{-0.63})$. This measurement becomes possible because the two redshift distributions (L2 and S1) nearly overlap and the signal is substantially influenced by the radial alignment of the S1 galaxies to the L1 galaxies.
    }
\label{fig:A_I_constraint}
\end{figure}

In our main presentation of the cosmological parameter estimation, we exclude the L2-S1 pair whose signal is substantially influenced by intrinsic alignments. We make this deliberate choice because we want to minimize the impact of the employed intrinsic alignment model, which we consider is incomplete.With the exclusion of the L2-S1 pair, we do not obtain any meaningful constraint on $A_{IA}$.
Here we present our results on the intrinsic alignment measurement when the L2-S1 pair is included in our parameter estimation. As expected, we obtain a significant constraint on $A_{IA}=2.51^{+0.82}_{-0.63}$ as shown in Figure\ref{fig:A_I_constraint}. This measurement becomes possible because the two redshift distributions (L2 and S1) nearly overlap and the signal is substantially influenced by the radial alignment of the S1 galaxies to the L1 galaxies. With Equation~\ref{eqn_IA} we find that the amplitude of the IA signal is roughly $\mytilde40$\% of the
L2-S1 $P^{gm}$ power spectrum for $A_{IA}=1$ (the sign is negative).

Our measurement $A_{IA} = 2.51^{+0.82}_{-0.63}$ is roughly consistent with the recent KIDS+2dFLenS \citep[$A_{IA} = 1.69\pm0.48$;][]{Joudaki:2017zdt} and KIDS+GAMA \citep[$A_{IA}=1.27\pm0.39$;][]{doi:10.1093/mnras/sty551} results. The statistical significance of the $A_{IA}$ parameter being positive is $\mytilde4\sigma$ in our case. 
The resulting $S_8$ value increases slightly to $0.829^{+0.034}_{-0.036}$
from our main result $0.810_{-0.031}^{+0.039}$ obtained without the L2-S1 pair. With the statistical errors considered, the two results are highly consistent with each other.

\label{sec:appendixf}
\begin{figure}
\centering
    \includegraphics[width = 0.64\textwidth]{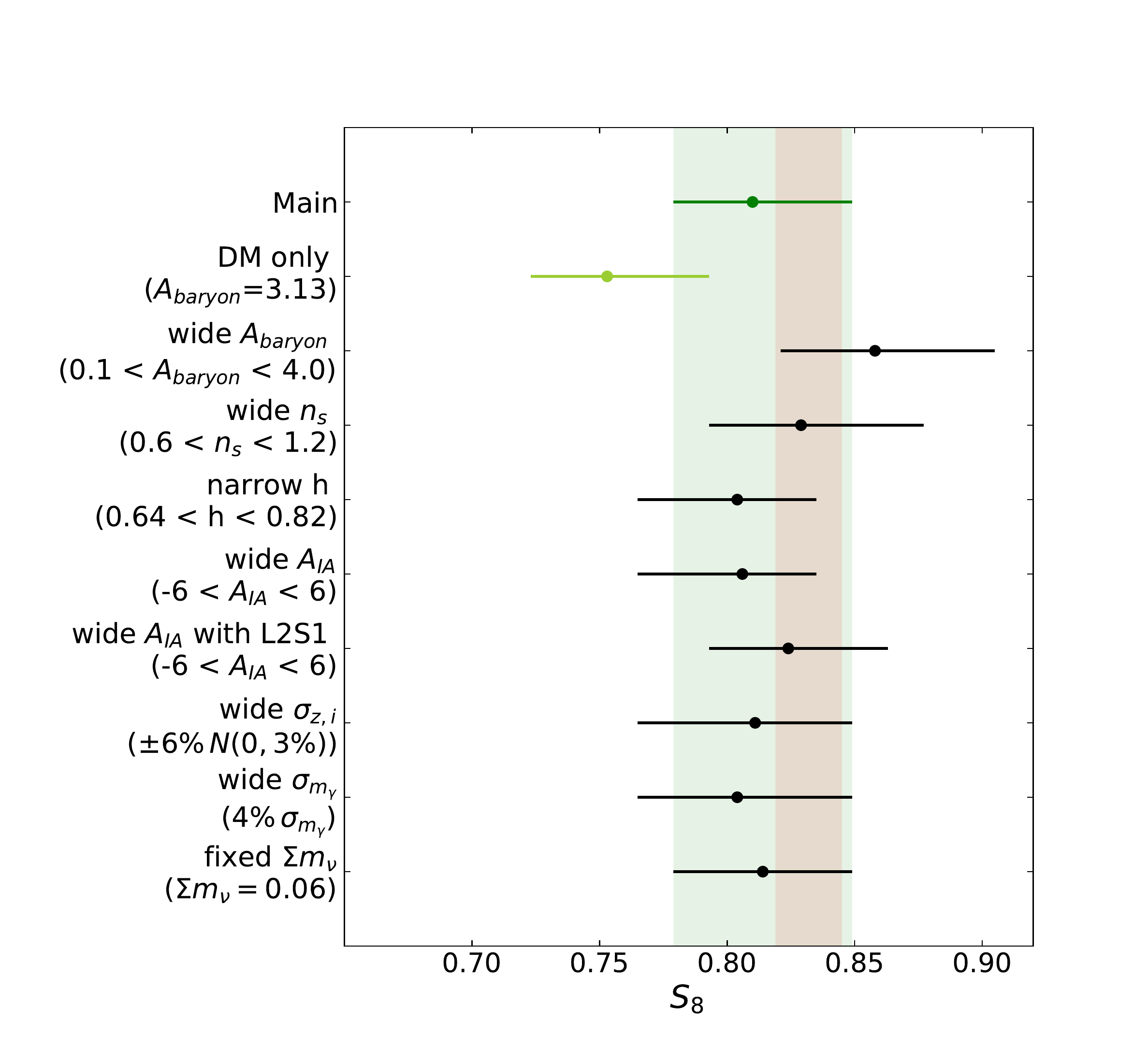}
    \caption{Impact of priors on $S_8$ constraints. The error bars are 1-$\sigma$ ranges. The green shaded region represents the constraint from the main result while the orange shaded region the Planck2018 constraint. Except for the dark matter only case, the test results are consistent with the Planck2018 constraint. See text for the description of each test label.}
\label{fig:s8_priors}
\end{figure}

\section{Appendix F. Impact of priors on $S_8$ }

Surveys with limited statistical powers result in cosmological parameter constraints that depend on imposed prior choices and their ranges. This causes a difficulty when results from different surveys and methods are compared \citep{Chang:2018rxd}.
We have tested impacts of prior ranges on the $S_8$ constraints for different choices of the baryonic feedback parameter $A_{baryon}$, power spectrum spectral index $n_s$, Hubble constant $h$, intrinsic alignment amplitude $A_{IA}$, photometric redshift systematics marginalization parameter $\sigma_z$, multiplicative shear calibration bias $\sigma_{m_\gamma}$, and sum of neutrino masses $\sum m_{\nu}$. Also, we examine the case when the L2-S1 pair is included in our cosmological parameter estimation.  

The test results are summarized in Figure~\ref{fig:s8_priors}, which shows that the $S_8$ values obtained from all test cases are consistent with one another, except for the case ``DM only" where we fix the baryonic feedback parameter to $A_{baryon}=3.13$, which corresponds to the OWLS DM-only simulation. This ``DM only" case still has overlapping error bars with most results (a slight tension exists between this ``DM only" and the ``wide $A_{baryon}$" cases) and is the only one that possesses a slight tension with the Planck2015 result.

The ``wide $A_{baryon}$" test refers to the case when we extend the $A_{baryon}$ prior interval to $[0.1,4.0]$. As mentioned in \S\ref{sec:baryon_model_sel}, the result favors $A_{baryon}\lesssim2$, which is the regime that has not been validated with numerical simulations. The increase in $S_8$ is consistent with our expectation because the lower $A_{baryon}$ value implies a higher power spectrum suppression.

For the other parameters, variation of priors yields very minor changes in $S_8$. Neither the ``wide $n_s$" test with $0.6 <n_s< 1.2$ ( $[0.86, 1.05]$ in the main setting) nor the ``narrow $h$" test with $0.65 <h< 0.82$ ($[0.55, 0.85]$ in the main setting) does not produce any significant shift.
The ``wide $A_{IA}$" test with $-6<A_{IA}<6$ ($[-4,4]$ in the main setting) does not degrade our constraining power on $S_8$ and only shifts the central value by $\mytilde0.004$. When we include the L2-S1 pair, the shift in $S_8$ is only $\mytilde0.014$ (``wide $A_{IA}$ with L2S1").
The assumption of larger shear multiplicative (4\%) and photometric redshift ($\pm6$\%) systematic errors ( referred to as ``wide $\sigma_{z,i}$" and ``wide $\sigma_{m_{\gamma}}$", respectively) lead to a $\mytilde20$\% increase in the uncertainty of $S_8$ with negligible changes in the central values.
Fixing the sum of neutrino masses to 0.06 (``fixed $\Sigma m_{\nu}$") gives the result that precisely overlaps with the main result.

\section{Appendix G. Scale dependence of baryonic feedback constraint}

\begin{figure}
    \centering
    \includegraphics[width = 0.45\textwidth]{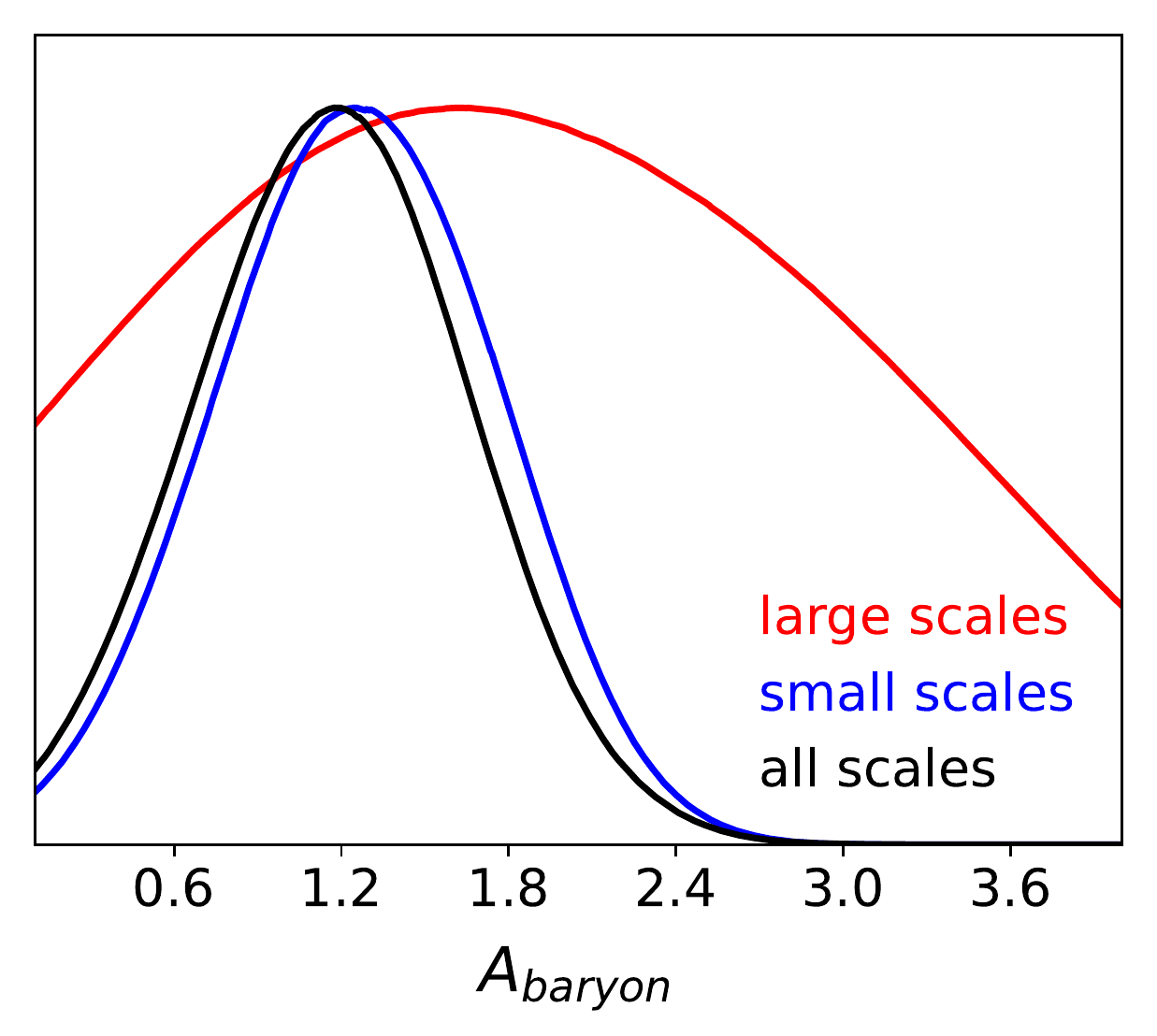}
    \caption{Scale dependence of baryonic feedback ($A_{baryon}$) constraint. Independent constraints on baryonic feedback from small scales (blue) and from large scales (red) are consistent with the main constraint obtained from all scales (black). The constraining power comes from nearly all scales, although the contribution is certainly dominated by the signals on small scales.}
    \label{fig:A_baryon_scale_dependence}
\end{figure}

The power spectrum suppression due to baryonic feedback increases for decreasing scales. Here we provide
consistency checks by repeating the measurement of $A_{baryon}$ using DLS signals (without the Planck2015 data) on different scales.
We fix the other parameters and their priors to the same values in the main setting.
The result is shown in Figure~\ref{fig:A_baryon_scale_dependence}. The blue solid curve shows the result $A_{baryon}=1.28^{+0.48}_{-0.45}$ when we use only the three largest $\ell$ bins (three smallest scales). This result is in good agreement with our main result $A_{baryon} = 1.19^{+0.51}_{-0.45}$ based on all five $\ell$ bins.
The red solid curve is obtained when we use only the two smallest $\ell$ bins (two largest scales). The constraint is weaker ($A_{baryon} =1.91^{+1.18}_{-1.44}$), but is still consistent with the main result. This experiment illustrates that the constraining power on $A_{baryon}$ comes from nearly all scales, although the contribution is certainly dominated by the signals on small scales.

\end{document}